\documentstyle[11pt,a4,cite,epsf]{article}
\newcommand{\lesssim}{ {\
\lower-1.2pt\vbox{\hbox{\rlap{$<$}\lower5pt\vbox{\hbox{$\sim$}}}}\ } }
\newcommand{\gtrsim}{ {\
\lower-1.2pt\vbox{\hbox{\rlap{$>$}\lower5pt\vbox{\hbox{$\sim$}}}}\ } }

\newcommand{\be}{\begin{equation}}
\newcommand{\ee}{\end{equation}}
\newcommand{\bea}{\begin{eqnarray}}
\newcommand{\eea}{\end{eqnarray}}
\newcommand{\noi}{\noindent}
\newcommand{\nn}{\nonumber}

\newcommand{\cL}{{\cal L}}
\newcommand{\cO}{{\cal O}}

\newcommand{\cA}{{\cal A}}

\newcommand{\Imm}{\mbox{\rm Im}}

\newcommand{\tr}{\mbox{\rm tr}}

\newcommand{\MeV}{\mbox{\rm MeV}}
\newcommand{\GeV}{\mbox{\rm GeV}}

\newcommand{\with}{\mbox{\rm with}}
\newcommand{\annd}{\mbox{\rm and}}

\newcommand{\al}{\alpha}
\newcommand{\als}{\alpha_{\mbox{\rm {\scriptsize s}}}}

\newcommand{\Lac}{\Lambda_{\chi}}
\newcommand{\Lmsb}{\Lambda_{\overline{\mbox{\rm\footnotesize MS}}}}

\newcommand{\res}{\mbox{\rm {\scriptsize res}}}

\newcommand{\Pivbar}{\bar{\Pi}}
\newcommand{\Pia}{\Pi_A}

\newcommand{\QCD}{QCD$(\infty)\;$}

\newcommand{\m}{\mu}
\newcommand{\n}{\nu}
\newcommand{\N}{N_c}

\newcommand{\eff}{\mbox{\rm{\tiny eff}}}

\newcommand{\EM}{\mbox{\tiny\rm  EM}}
\newcommand{\ENJL}{\mbox{\rm{\tiny ENJL}}}

\newcommand{\ra}{\rightarrow}

\input epsf

\begin{document}
\begin{titlepage}

\begin{flushright} CPT-97/P.3638\\ UAB-FT-443\\ 


\end{flushright}
\vspace*{1.5cm}
\begin{center} {\Large \bf Matching Long and Short Distances
             \\[0.5cm] in Large-$N_c$ QCD}\\[3.0cm] {\large {\bf
Santiago Peris}$^{a}$, {\bf Michel Perrottet}$^b$ and {\bf
Eduardo de Rafael}$^b$}\\[1cm]

$^a$ Grup de F{\'\i}sica Te{\`o}rica and IFAE\\ Universitat
Aut{\`o}noma de Barcelona, 08193 Barcelona, Spain.\\[0.5cm]
and\\[0.5cm]
$^b$  Centre  de Physique Th{\'e}orique\\
       CNRS-Luminy, Case 907\\
    F-13288 Marseille Cedex 9, France\\

\vskip 1cm

\sl{Submitted to JHEP}

\end{center}

\vspace*{1.0cm}

\begin{abstract}

It is shown, with the example of the experimentally known  Adler function,
that there is no matching in the intermediate region between the two
asymptotic regimes described by perturbative QCD (for the very
short--distances) and by chiral perturbation theory (for the very
long--distances).  We then propose to consider an approximation of
large--$N_c$ QCD which consists in restricting the hadronic spectrum in the
channels with $J^P$ quantum numbers
$0^{-}$, $1^{-}$, $0^{+}$ and $1^{+}$  to the lightest
state and to treat the rest of the narrow states as a
perturbative QCD continuum;  the onset of this continuum being fixed by
consistency constraints from the operator product expansion. We show how to
construct the low--energy effective Lagrangian which describes this
approximation. The number of free parameters in the resulting effective
Lagrangian can be reduced, in the chiral limit where the light quark masses
are set to zero, to just one mass scale and one dimensionless constant to all
orders in chiral perturbation theory. A comparison of the corresponding
predictions, to
$\cO(p^4)$ in the chiral expansion, with the phenomenologically known
couplings is also made.    
\end{abstract}
\end{titlepage}

\section{Introduction}
\label{sec:Introduction}

We shall be concerned with Green's functions of colour singlet operators
which are bilinear in the light quark fields. There are at present two
asymptotic regimes where QCD is a predictive theory for these Green's
functions: the short--distance regime where the external momenta are deep
euclidean in regions of non--exceptional momenta~\cite{Sy71},  and the
very long--distance regime where the external momenta are very soft. The
behaviour of the Green's functions in the short--distance regime can be
successfully predicted via perturbation theory in powers of the QCD
running coupling constant (pQCD). In the chiral limit where the light
quark masses are set to zero, the behaviour in the short--distance regime
is then governed by only one mass scale, e.g. the $\Lmsb$ parameter for three
flavours.

The way that QCD predicts the behaviour of the Green's functions in the
long--distance regime is not so straightforward. At very low energies, the
sole input of the property of spontaneous breakdown of the chiral
$SU_L(3)\times SU(3)_R$ invariance of the QCD Lagrangian with
massless $u$, $d$, $s$ quarks, allows
one to set up a well defined theoretical
framework~\cite{We67,We79,GL84,GL85,L94} known as Chiral Perturbation Theory
($\chi$PT). This is formulated in terms of an effective Lagrangian that
describes, at low energies, the strong interactions of the Goldstone
modes associated with spontaneous chiral symmetry breaking. The behaviour
of the Green's functions in the very long--distance regime is then
predicted by this effective Lagrangian  as an expansion in
powers of momenta. This, however, comes at the expense of introducing an
increasing number of local operators of higher dimension with unknown
couplings which build the effective Lagrangian at higher orders
in the chiral expansion. [Some of these local operators act as
counterterms in the loop expansion generated by the lower order terms.]
At
$\cO(p^6)$ in the chiral expansion the number of independent local
operators is already of the order of a hundred~\cite{FS96}. Although in
principle their couplings are fixed from QCD, their calculation demands a
genuine knowledge of non--perturbative QCD properties. In certain
circumstances one can relate these low--energy constants to a sufficient
amount of independent experimental observables which determine them and many
successful predictions have been made this way~\footnote{See e.g. the
review articles in refs.~\cite{Pic95,deR95,Eck95}}. [This is the orthodox
methodology of the
$\chi$PT science.] However, as higher orders in the chiral expansion are
required, it often happens that there is simply no model independent way
to fix all the couplings. In practice the
onset of this situation is at
$\cO(p^6)$ in the chiral Lagrangian for the strong interactions and
already at
$\cO(p^4)$ in the sector of the non--leptonic weak
interactions.  If one wishes to make
progress here, one clearly needs to find a good theoretical approximation
as a substitute for full fledged QCD.

An interesting question one may ask is: what is the {\it matching} of the
two asymptotic regimes which we have just described? This is precisely the
question which we wish to discuss in this paper. For simplicity we shall
illustrate most of this discussion with a specific Green's function,
the so called Adler function (see section 2 below for definitions), but
many of the features presented here are rather generic. We claim that the
direct  matching is in fact extremely poor; even when non--perturbative
short--distance power corrections
{\`a} la SVZ~\cite{SVZ79} are retained. Once more we find that if one
wishes to make progress in the quality of the  matching between the
pQCD short--distance behaviour and the $\chi$PT behaviour which is
only operational at very long--distances, one clearly needs a good
theoretical approximation as a substitute for full fledged QCD.

As a first step in this direction we shall consider QCD in the
limit of a large number of colours~\cite{'tH74,RV77,W79,Wi80,Ma98}.
This limit, which following 't Hooft's notation  we shall term
\QCD for short, offers a convenient starting point where one can hope to
discuss fruitfully the aforementioned matters since it is still rich
enough to share with QCD its most relevant non--perturbative features.
Among them, a very important one, is the fact that \QCD has the correct
pattern of spontaneous chiral symmetry breaking~\cite{CW80,KdeR97},
provided that the property of confinement still persists in going from
QCD to QCD($\infty$). The advantage of the large$-\N$ limit is the
emergence of narrow meson states in the Green's functions. The
disadvantage is that the number of narrow states is infinite and they
appear at all energy scales, even when the euclidean momenta are large
enough that a perturbative treatment should be adequate. Obviously, a
perturbative description, when possible, is more predictive than an
infinite tower of narrow states with masses and decay constants which are
a priori unknown.

We propose to consider an approximation to \QCD which consists
in restricting the hadronic spectrum in the channels with $J^P$ quantum
numbers $0^{-}$, $1^{-}$, $0^{+}$ and $1^{+}$  to the lowest energy state and
in treating the rest of the narrow states as a
\QCD perturbative continuum; 
the onset of the continuum being fixed by consistency
constraints from the operator product expansion (OPE) as we shall discuss.
The reason why we limit ourselves to this simple hadronic spectrum
is that, in the large--$N_c$ limit, this is the minimal structure compatible
with spontaneous chiral symmetry breaking~\cite{KdeR97}. The traditional
successes of ``vector meson dominance'' in describing low energy hadron
phenomenology guarantees that this approximation, which we shall call
``lowest meson dominance (LMD) approximation to large--$N_c$ QCD'',  should
be rather reasonable~\footnote{If necessary, one could also add further
states, (see e.g. ref.~\cite{AET98} for a recent
discussion of a possible description of the $\pi(1300)$.) However, as we
shall see, the phenomenological success of the LMD approximation indicates
that, to a good approximation, these higher states play a minor r{\^o}le in
the dynamics of the low--energy effective Lagrangian.}.  The immediate
question which then arises is the following: is there an effective Lagrangian
formulation  which explicitly exhibits the properties of
\QCD in the LMD approximation? We shall show that there is indeed such an
effective Lagrangian. The effective Lagrangian in question can be obtained
starting from a modified version of an extended Nambu--Jona-Lasinio type
Lagrangian (ENJL) which has been very much discussed in the
literature~\footnote{See e.g. refs.~\cite{NJL61,DSW85,BBdeR93,BdeRZ94,Bi96}
and references therein}. The modifications, however, are highly non trivial
because they require the presence of an infinity of local operators which,
however, have  known couplings. The couplings of the new operators are known
because they are fixed by the requirement that they  remove, order by order
in the chiral expansion, the effects of  {\it non--confining}
$Q\bar{Q}$ discontinuities produced by the initial ENJL {\it ansatz}. 
The resulting Lagrangian when the quark fields are integrated out is
a non--linear sigma type Lagrangian with degrees of freedom
corresponding to the Goldstone fields and also to Vector, Axial--Vector, and
Scalar fields. The {\it matching} of the two--point functions of this
effective Lagrangian to their QCD short--distance behaviour can be
systematically implemented. However, for Green's functions beyond two--point
functions, the removal of the {\it non--confining}
$Q\bar{Q}$ discontinuities produced by the initial ENJL {\it ansatz} is not
enough to guarantee in general the correct {\it matching} to the leading QCD
short--distance behaviour and further local operators have to be included. 
In the sector of Vector and Axial--Vector couplings, and
after the {\it matching} with the QCD short--distance behaviour from the OPE
for three--point functions is implemented~\cite{MOU97}, the resulting
low--energy Lagrangian to
$\cO(p^4)$ in the chiral expansion coincides with the class of Lagrangians
discussed in ref.~\cite{EGLPR89}. The advantage, however, with respect to a
purely phenomenological description in terms of chiral effective Lagrangians
which include resonances as discussed e.g. in
refs.~\cite{EGPR89,DRV89,DHW93,DP97} and references therein, is that in the
LMD approximation to
\QCD, the number of free parameters can be reduced to just one mass
scale and one dimensionless parameter to all orders in
$\chi$PT.

\section{The Adler Function}
\setcounter{equation}{0}
\label{sec:two}

\noi
We are concerned with two-point functions
\be\label{eq:tpfV}
\Pi_{\mu\nu}(q)_{ab}=i\int d^{4}xe^{iq\cdot x}<0|T\{V_\mu^{a}(x)
V_\nu^{b}(0)\}|0>
\ee
of vector quark currents
\be\label{eq:Vcur}  V_\mu^{a}(x)=\bar q (x)\gamma_\mu
\frac{\lambda^{a}}{\sqrt{2}}q(x)\,,
\ee
where $\lambda^{a}$ are Gell-Mann matrices
($\tr\lambda^{a}\lambda^{b}=2\delta_{ab}$) acting on the flavour triplet
of $u$, $d$, $s$ light quarks. In the chiral limit where the
light quark masses are set to zero, these two--point functions depend only
on one invariant function ($Q^2 =-q^2\ge 0$ for $q^2$ spacelike)
\be\label{eq:itpfV}
\Pi_{\mu\nu}(q)_{ab}=(q_\mu q_\nu -g_{\mu\nu}q^2)
\Pi(Q^2)\delta_{ab}\,.
\ee
In QCD, the invariant function $\Pi(Q^2)$ obeys a once--subtracted
dispersion relation and it is conventional to choose the subtraction at
$Q^2=0$:
\be\label{eq:dr}
\Pi(Q^2)=\Pi(0)-Q^2\int_0 ^{\infty}\frac{dt}{t}\frac{1}{t+Q^2}
\frac{1}{\pi}\,\Imm\Pi (t)\,.
\ee
The Adler function~\cite{Ad74} is defined as the 
derivative of the invariant function $\Pi(Q^2)$ i.e.,
\be\label{eq:AdlerV}
\cA (Q^2) = -Q^2\frac{d\Pi(Q^2)}{d Q^2}\,,
\ee
and therefore, its relation to the spectral function
$\frac{1}{\pi}\Imm\Pi(t)$ does not depend on the choice of the subtraction
\be\label{eq:Pdr}
\cA (Q^2)=\int_0 ^{\infty}dt\frac{Q^2}{(t+Q^2)^2}\frac{1}{\pi}\,\Imm\Pi
(t)\,.
\ee

\vspace*{0.5cm}

\subsection{Short--Distance Behaviour}
\label{subsec:twoone}

\noi The Adler function obeys a homogeneous renormalization group
equation, and it has been computed in pQCD (in the
$\overline{MS}$ renormalization scheme) up to ${\cal O}(\alpha_s^3)$, with
the result~\cite{GKL91,SS91}
\bea\label{eq:PQCD}
\cA (Q^2)\vert_{\mbox{\rm pQCD}}=
\frac{N_c}{16\pi^2}\frac{4}{3}\left\{ 1 + \frac{\alpha_s
(\mu^2)}{\pi}+\left[F_2 +
\frac{\beta_1}{2}\log{\frac{Q^2}{\mu^2}}\right]\left(\frac{\alpha_s
(\mu^2)}{\pi}\right)^2 \right.  \nonumber \\ +\left.\left[F_3 +
(F_2\beta_1+\frac{\beta_2}{2})\log{\frac{Q^2}{\mu^2}}+\frac{\beta_1^2}{4}
(\frac{\pi^2}{3}+{\log}^2{\frac{Q^2}{\mu^2}})\right] \left(\frac{\alpha_s
(\mu^2)}{\pi}\right)^3 + {\cal O} (\alpha_s^4)\right\}\,,
\eea
where
\be\label{eq:betas}
\beta_1 = \frac{n_f}{3}-\frac{11}{6}N_c\,,  \qquad
\beta_2 = -\frac{51}{4}+\frac{19}{12}n_f\,,
\ee
\be
 F_2 = 1.986 - 0.115\, n_f \,,\qquad
 F_3 = -6.637 - 1.200\, n_f - 0.005\, n_f^2\,,
\ee
and $n_f$ the number of light quark flavors. The behaviour of the Adler
function as a function of $Q^2$ predicted by  pQCD is shown in Fig.~1. The
plot in Fig.~1 corresponds to the choice $\mu^2=Q^2$ in
eq.~(\ref{eq:PQCD}),  with
$\alpha_s (Q^2)$  the solution to the implicit equation
\be\label{eq:alphas}
\frac{2\pi}{-\beta_1 \alpha_s(Q^2)}=\left(\log
\frac{Q^2}{\Lambda^2}\right)
\left( 1 - \frac{2\beta_2}{\beta_1^2\log \frac{\mu^2}{\Lambda^2}} \log
\left[
\frac{2\pi}{-\beta_1 \alpha_s(Q^2)} - \frac{2\beta_2} {\beta_1^2}
\right] \right)\,,
\ee
which results from solving the running coupling constant
renormalization group equation at the two--loop level. Figure~1 clearly
shows the asymptotic freedom behaviour:
$\cA (Q^2\ra\infty)=
\frac{N_c}{16\pi^2}\frac{4}{3}$, as well as the fact that at $Q^2\lesssim
1\GeV^2$ we are already entering a region where non--perturbative effects
are likely to be important.

\vspace{0.3cm}
\centerline{\epsfbox{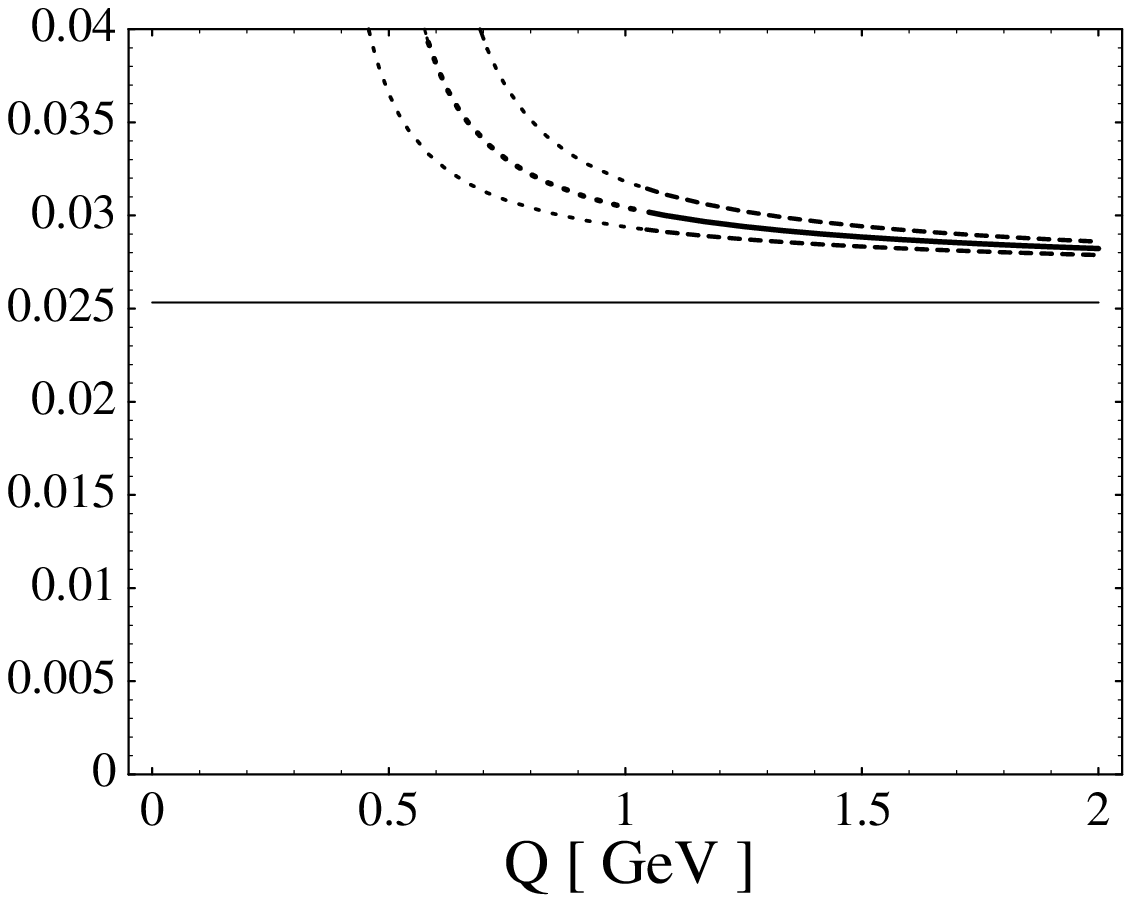}}
\vspace{0.2cm}

{\bf{Fig.~1}}
{\it{The Adler function from perturbative QCD. The dashed lines
correspond\newline to
$\Lambda_{\overline{MS}}= (372 \pm 76)\,\MeV$}}~\cite{lambda1,lambda2}.

\vspace{0.3cm}

As first pointed out by Shifman, Vainshtein and Zakharov~\cite{SVZ79},
non--perturbative
$\frac{1}{Q^2}$--power corrections appear naturally when two--point
functions like eq.~(\ref{eq:tpfV}) are considered in the physical vacuum
within the framework of the operator product expansion (OPE). The leading
$\frac{1}{Q^4}$ and next--to--leading $\frac{1}{Q^6}$ non--perturbative
corrections were calculated in ref.~\cite{SVZ79}. They depend on the size
of the {\it gluon condensate}, which gives the leading correction to the
perturbative result in eq.~(\ref{eq:PQCD}):
\be
\cA (Q^2)_{GG}=\frac{1}{6}\frac{\als}{\pi}\frac{\langle
G_{a}^{\mu\nu}G^{a}_{\mu\nu}\rangle}{Q^4}\,;
\ee
and on the size of  {\it four quark condensates}, which give the
next--to--leading power correction to the perturbative result in
eq.~(\ref{eq:PQCD}):
\bea
\cA (Q^2)_{{\bar\psi}\psi{\bar\psi}\psi}
=\left[-3\left\langle\left( {\bar
u}\gamma_{\mu}\gamma_{5}\lambda^{a}u- {\bar
d}\gamma_{\mu}\gamma_{5}\lambda^{a}d\right)^2\right\rangle\right. \nn \\
-\left.
\frac{2}{3}\left\langle\left( {\bar u}\gamma_{\mu}\lambda^{a}u+{\bar
d}\gamma_{\mu}\lambda^{a}d\right)\sum_{q=u,d,s}{\bar
q}\gamma^{\mu}\lambda^{a}q\right\rangle\right]
\frac{\pi\als}{Q^6}.
\eea
To leading order in the $1/N_c$--expansion {\it four quark
condensates} factorize into products of the lowest dimension {\it quark
condensate} with the result
\be
\cA (Q^2)_{{\bar\psi}\psi{\bar\psi}\psi}^{N_c\ra\infty}=
-\frac{28}{3}\frac{\pi\als}{Q^6}\langle\bar{\psi}\psi\rangle^2\,.
\ee
\noi
The phenomenological determinations of the {\it gluon condensate} and the
{\it quark condensate} have unfortunately rather big errors. A generous
range for the {\it gluon condensate} which covers most of the
determinations is
\be\label{eq:gluonc}
\langle\als G_{a}^{\mu\nu}G^{a}_{\mu\nu}\rangle=(0.08\pm 0.04)\,\GeV^4.
\ee
The {\it quark condensate} is scale dependent. The values obtained
from a variety of sum rules~\footnote{See e.g. the recent compilation of sum
rule estimates in ref.~\cite{DN97}. Our error in eq.~(\ref{eq:quarkc}) is
however larger than the one quoted in this reference.}
\be\label{eq:quarkc}
\langle\bar{\psi}\psi\rangle(1\,\GeV^2)=
[(-235\pm 27)\,\MeV]^3\,.
\ee
Using these values, the resulting prediction for the Adler function is
shown in Fig.~2.

\vspace{0.3cm}
\centerline{\epsfbox{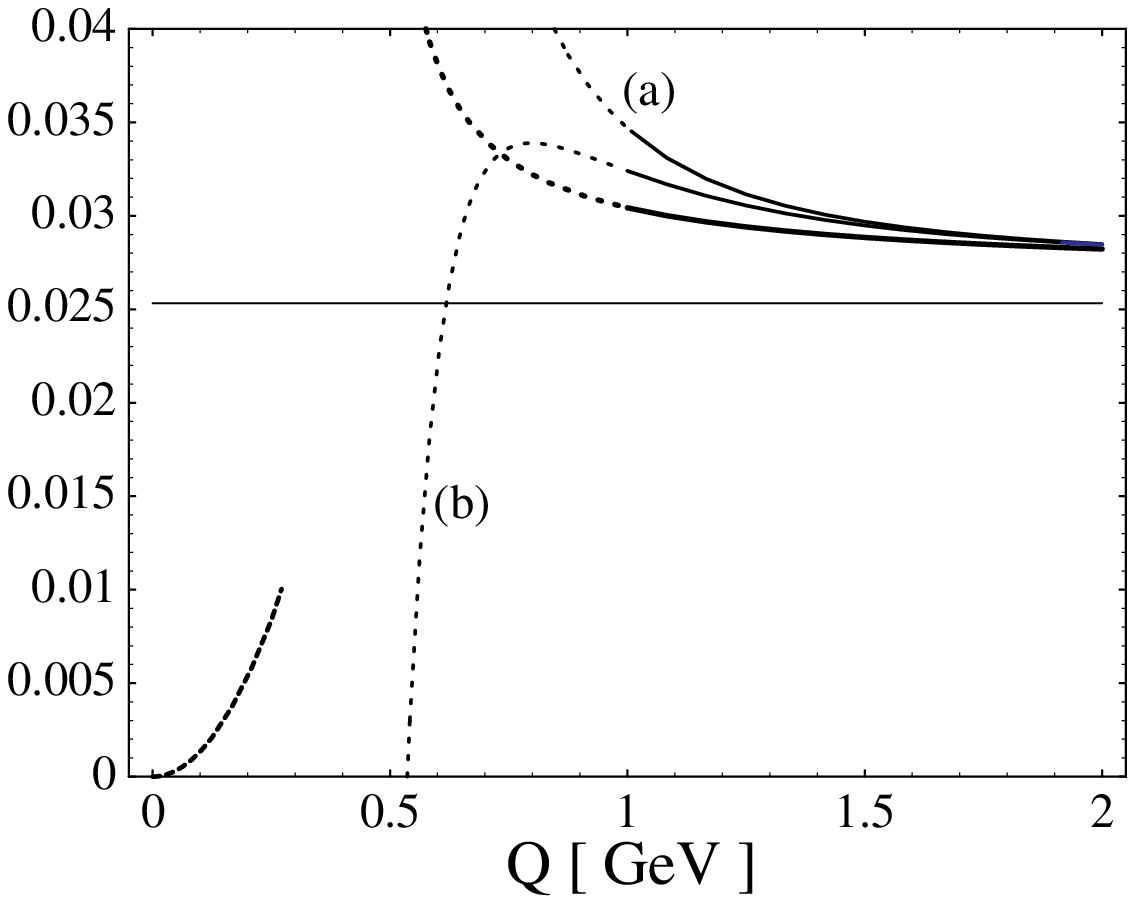}}
\vspace{0.2cm}

{\bf{Fig.~2}}
{\it{ The Adler function including (a) the $\frac{1}{Q^4}$ corrections, 
(b) the $\frac{1}{Q^4}$  and $\frac{1}{Q^6}$ corrections. The dashed line is
the VMD prediction corresponding to eq.~(\ref{eq:VMDph}) below, valid at low
$Q$.}}

\vspace{0.3cm}
\noindent
 The curve $(a)$ in the figure is the one which
results from adding only the $\frac{1}{Q^4}$--corrections to the pQCD
result shown in Fig.~1; the  curve $(b)$ includes both the
$\frac{1}{Q^4}$ and $\frac{1}{Q^6}$--corrections.  The curves in Fig.~2
clearly show that with the values for the lower dimension {\it
condensates} given in eqs.~(\ref{eq:gluonc}) and (\ref{eq:quarkc}), the
contribution from the leading and next--to--leading non--perturbative
power corrections -at $Q^2$ values where one can trust the pQCD
contributions-  turn out to be already quite small. This is due to the
empirical fact that the scale of perturbation theory
$\Lambda_{\overline{MS}}$ is rather large (we are
using~\cite{lambda1,lambda2}
$\Lambda_{\overline{MS}}=(372\pm 76)\,\MeV$ in our plots). For values of
$\Lambda_{\overline{MS}}$ as large as that, one is forced to choose
$Q^2>1\,\GeV^2$ to trust the perturbative expansion in powers of
$\als(Q^2)$, and for $Q^2>1\,\GeV^2$ the power
corrections are already quite small.

\vspace*{0.5cm}

\subsection{Long--Distance Behaviour}
\label{subsec:twotwo}

\noi
The behaviour of the invariant function $\Pi(Q^2)$ in eq.~(\ref{eq:itpfV})
at
$Q^2=0$ is governed by a combination of two $\cO (p^4)$ coupling
constants of the
$\chi$PT effective Lagrangian:
\be
\Pi(Q^2) = -4(2H_{1}+L_{10})+\cO (Q^2)\,.
\ee
The constants $H_{1}$ and $L_{10}$ are the coupling constants of the
terms
\bea
\cL_{\eff}(x) & = & \cdots + L_{10}\tr\,
U(x)^{\dagger}F_{R}^{\mu\nu}(x)U(x) F_{L\mu\nu}(x) \nn\\
 & & +H_{1}\tr\left(F_{R}^{\mu\nu}(x)F_{R\mu\nu}(x)+
F_{L}^{\mu\nu}(x)F_{L\mu\nu}(x)\right)+\cdots
\eea
in the effective chiral Lagrangian, where $F_{L}$ and $F_{R}$ are the
field strength tensors associated with external $v_{\mu}-a_{\mu}$ and
$v_{\mu}+a_{\mu}$ matrix field sources, and $U(x)$ the unitary matrix which
collects the Goldstone
fields ($U\ra V_{R}UV_{L}^{\dagger}$ under chiral transformations). The Adler
function is not sensitive to
$\Pi(0)$, and hence to $2H_{1}+L_{10}$. The slope of the Adler function
at the origin corresponds therefore to combinations of couplings in
$\chi$PT which are already of $\cO (p^6)$ and therefore theoretically
unknown {\it a priori}. There are also contributions to
the slope of the Adler function from chiral loops involving lower order
couplings. These chiral loop contributions are calculable~\cite{GK95}, but
they are non--leading in the $1/N_c$--expansion and therefore we shall ignore
them here.

One can make an ``educated'' guess of the value of the slope of the Adler
function at the origin by invoking VMD arguments. In this case it amounts
to the assumption that, at low energies,  the vector spectral function is
dominated by the narrow width pole corresponding to the $\rho$--meson
\be
\frac{1}{\pi}\Imm\Pi(t)\simeq 2f_{\rho}^2 M_{\rho}^2 \delta (t-M_{\rho}^2)\,.
\ee
This leads to the ``prediction''
\be\label{eq:VMDph}
\cA(Q^2\ra 0)_{\mbox{\rm\scriptsize
VMD}}=\frac{2f_{\rho}^2}{M_{\rho}^2}Q^2 +\cO (Q^4)\,,
\ee
and it corresponds to the dashed line starting at $Q^2=0$ which we
have plotted in Fig.~2.

We conclude that the only rigorous knowledge we have from QCD about the
long--distance behaviour of the Adler function is that it vanishes at
$Q^2=0$, with a slope which is governed by $\cO (p^6)$ terms in $\chi$PT
and therefore its determination is, at present, model dependent. We wish
to emphasize, however, that the property that it vanishes at $Q^2=0$ is
highly non--trivial. There is nothing in the pQCD behaviour in
eq.~(\ref{eq:PQCD}) which points to that. Although one can find in the
literature arguments invoking that
$\als(Q^2)$ might ``freeze'' to a constant value when $Q^2\ra 0$, one
should be aware that in the case of the Adler function a na{\"\i}ve ``freezing
argument'' contradicts~\footnote{See the appendix in ref.~\cite{PdeR97} for
further discussion on this point.} the long--distance QCD behaviour encoded in
$\chi$PT. 
\vspace*{0.5cm}

\subsection{Empirical Determination of the Adler Function}
\label{subsec:twothree}

\noi
The isovector hadronic spectral function has been determined both
from $e^{+} e^{-}$--annihilation measurements and from hadronic
$\tau$--decays~\footnote{~For a recent review, where references to the
original experiments can be found, see ref.~\cite{Hoc97}.}. We show in
Fig.~3 the compilation from the $\tau$--data as given in
ref.~\cite{Hoc97}, except for the overall normalization.

\vspace{0.3cm}
\centerline{\epsfbox{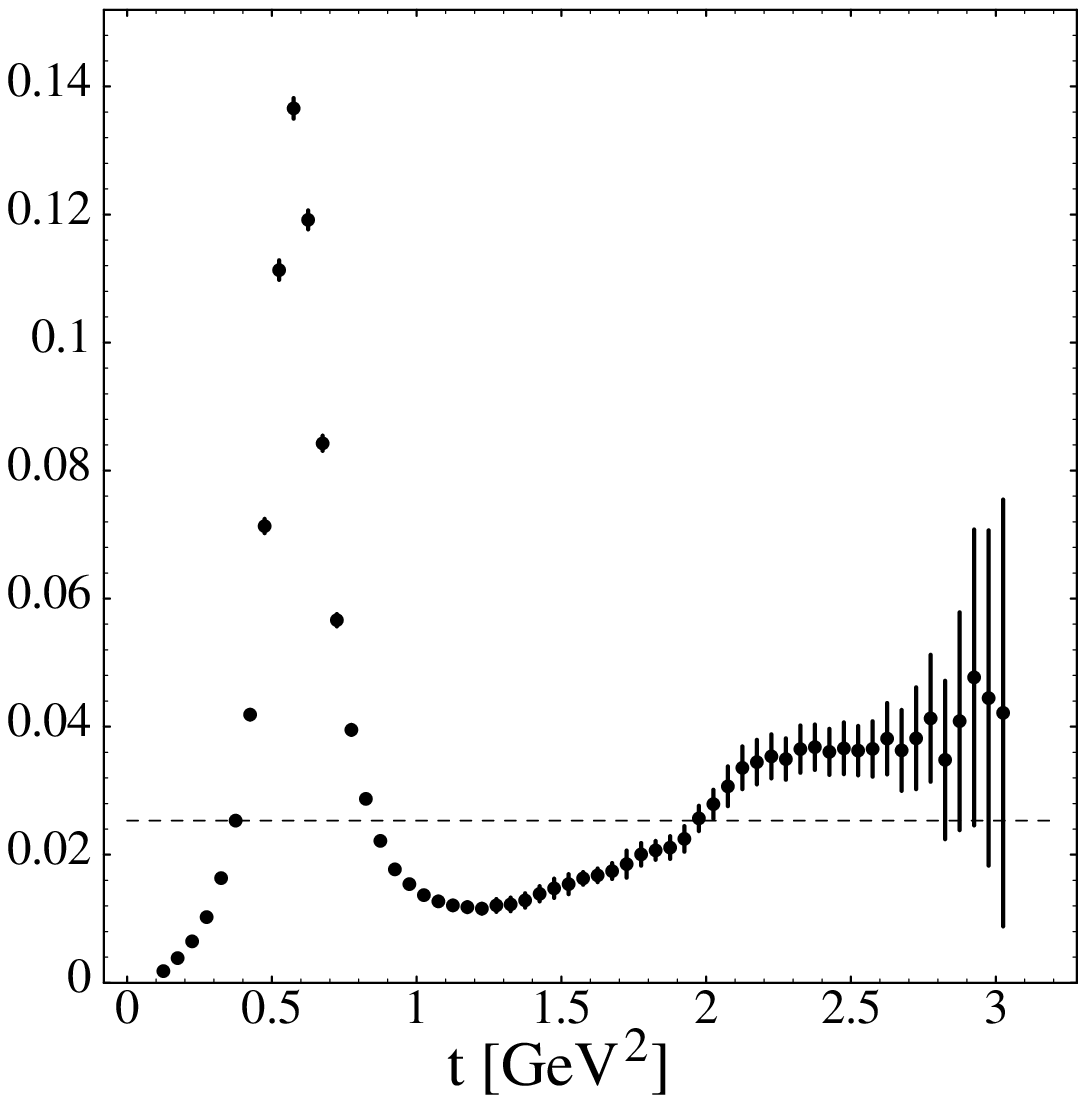}}
\vspace{0.2cm}

{\bf{Fig.~3}}
{\it{The isovector hadronic spectral function from hadronic $\tau$ - decays. 
The dashed line is the asymptotic freedom prediction $\frac{1}{4\pi^2}$.}}

\vspace{0.3cm}

\noindent
The normalization
in our Fig.~3 is in accordance with the definition of the spectral
function given in eqs.~(\ref{eq:tpfV}) to (\ref{eq:dr}) above. When
extracting the Adler function from experimental data, one is confronted
with the question of {\it matching} to the asymptotic QCD continuum beyond
the region accessible to experiment. The concept of {\it global duality}, as
explained in refs.~\cite{BLdeR85,deR98}, provides the way to do this {\it
matching}. In practice, it consists in introducing an {\it ansatz} for the
complete  spectral function
\be\label{eq:spectralansatz}
\frac{1}{\pi}\Imm\Pi(t)=\frac{1}{\pi}\Imm\Pi(t)_{\mbox{\rm\small
exp.}}\,\theta(s_{0}-t)+
\frac{1}{\pi}\Imm\Pi(t)_{\mbox{\rm\small pQCD}}\,\theta(t-s_{0})\,,
\ee
where $\frac{1}{\pi}\Imm\Pi(t)_{\mbox{\rm\small exp.}}$ denotes the
hadronic experimental spectral function and
$\frac{1}{\pi}\Imm\Pi(t)_{\mbox{\rm\small pQCD}}$ the spectral function
predicted by pQCD. The onset of the continuum
$s_{0}$ is then fixed by solving the equation in $s_{0}$:
\be\label{eq:eigenv}
\int_{0}^{s_{0}}dt\,\frac{1}{\pi}\Imm\Pi(t)_{\mbox{\rm\small exp.}}=
\int_{0}^{s_{0}}dt\,\frac{1}{\pi}\Imm\Pi(t)_{\mbox{\rm\small pQCD}}\,.
\ee
This equation guarantees that the {\it matching} between
long distances and short distances in the Adler function is consistent
with the OPE. Indeed, it can be easily shown that the equation
in (\ref{eq:eigenv}) follows from the absence of operators of dimension
two~\footnote{~In the presence of quark masses there is in fact a
correction term to eq.~(\ref{eq:eigenv}) of
$\cO(\frac{m^2}{s_{0}})$ which we are neglecting here.} in the OPE of the
product of current operators which defines the Adler function.

Using the experimental input corresponding to the $\tau$--decay data
shown in Fig.~3, and the expression~\footnote{~For the purposes we are
interested in here, the evaluation of this integral as a
contour integral in the complex t--plane, as proposed e.g. in
ref.~\cite{LP92}, does not change significantly the result.}
\bea\label{eq:experint}
\lefteqn{\int_{0}^{s_{0}}dt\,\frac{1}{\pi}\Imm\Pi(t)_{\mbox{\rm\small pQCD}}=
\frac{N_c}{16\pi^2}\frac{4}{3}s_{0}\left\{1+\frac{\als(s_{0})}{\pi}\right.}
\nn\\ & & \left. +\left[F_{2}-\frac{\beta_{1}}{2}\right]
\left(\frac{\als(s_{0})}{\pi}\right)^2+\left[F_{3}-
\left(F_{2}\beta_{1}+\frac{\beta_{2}}{2}\right)+\frac{\beta_{1}^2}{2}
\right]\left(\frac{\als(s_{0})}{\pi}\right)^{3}\right\}\,,
\eea
for the pQCD r.h.s. in eq.~(\ref{eq:eigenv}) results in the 
solution
\be s_{0}=(1.60\pm 0.17)\,\GeV^2 \, ,
\ee
where the error comes from the error on $\Lambda_{\overline{MS}}$ only.
The resulting Adler function is shown in Fig.~4 below.

\vspace{0.3cm}
\centerline{\epsfbox{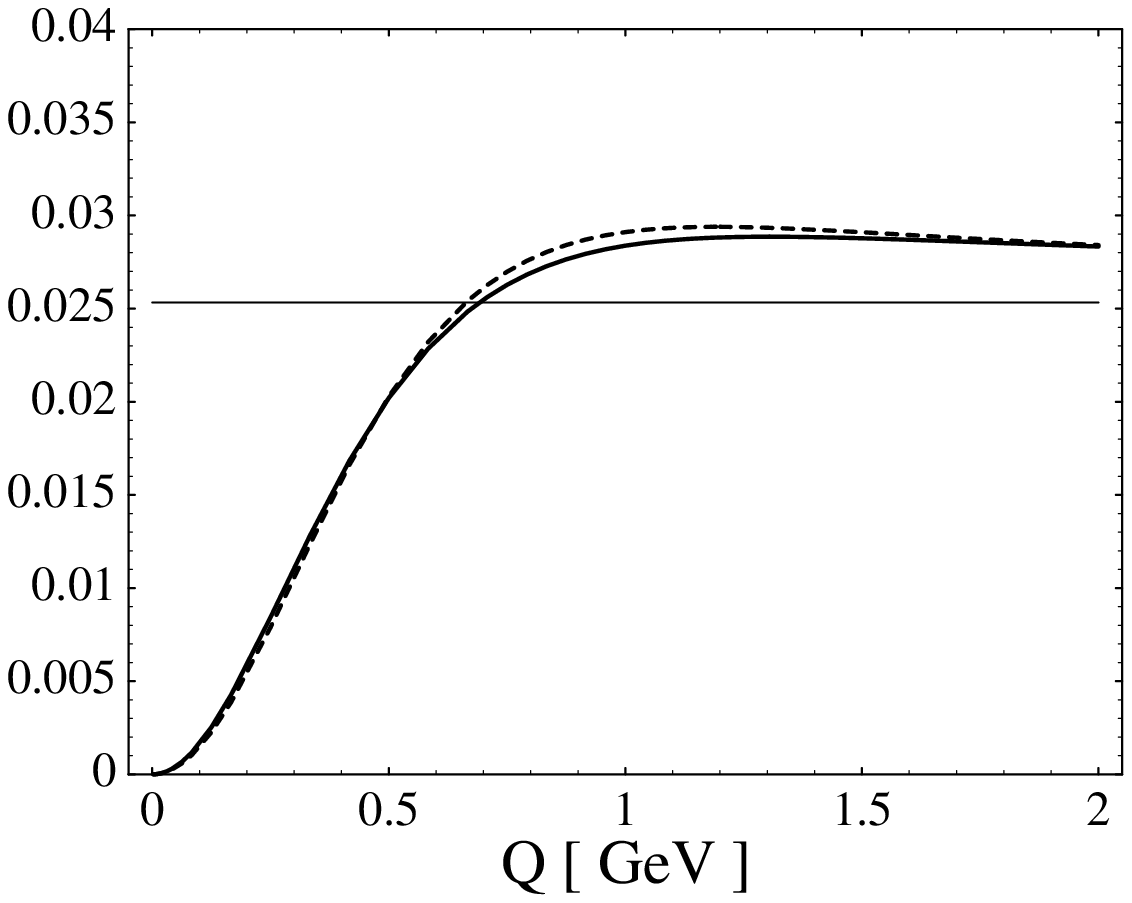}}
\vspace{0.2cm}

{\bf{Fig.~4}}
{\it{The Adler function from experimental data (full line) and from the VMD
parameterization (dashed line) in eq.(\ref{eq:VMD}). }}

\vspace{0.3cm}
\noindent
It is interesting to compare the ``full experimental'' result shown in
Fig.~4 with the one obtained from the simplest hadronic parameterization one
can think of, which is the one corresponding to a narrow width vector
state --the
$\rho(770)$-- \underline{plus} the corresponding pQCD continuum i.e.,
\be\label{eq:VMD}
\frac{1}{\pi}\Imm\Pi(t)=2f_{\rho}^2 M_{\rho}^2\delta(t-M_{\rho}^2)+
\frac{1}{\pi}\Imm\Pi(t)_{\mbox{\rm\small pQCD}}\,\theta(t-s_{\rho})\,.
\ee
The onset of the continuum for this parameterization is at 
$s_{\rho}=(1.53\pm 0.06)\,\GeV^2$, and the corresponding
Adler function is also shown in Fig.~4 -the dashed curve- . Although
point by point the two spectral functions in eqs.~(\ref{eq:spectralansatz})
and (\ref{eq:VMD}) are very different below their continuum onset, the
associated Adler functions result in rather similar functions of the
euclidean variable
$Q^2$. It is in this sense that these two spectral functions are 
{\it globally
dual}~\cite{BLdeR85,deR98}.

\vspace*{0.5cm}

We wish to summarize the basic features which we have learnt in this
section.
\begin{itemize}
\item The direct matching between $\chi$PT and pQCD is very poor, regardless
of whether or not one includes non--perturbative power corrections {\`a} la
SVZ~\cite{SVZ79}. Including these corrections does not help much because
of the empirical fact that $\Lmsb$ is large. At the $Q^2$ values above
which we can trust pQCD, the non--perturbative power corrections are
already quite small.

\item As shown in Fig.~2 there is no  matching between the
na{\"\i}ve extrapolation of the linear
$Q^2$ behaviour predicted by the first  non--trivial order of $\chi$PT, 
$\cO(p^6)$ in this case, and the pQCD prediction with
inclusion of the leading
$\frac{1}{Q^4}$ and next--to--leading $\frac{1}{Q^6}$ non--perturbative
power corrections.

\item The simplest hadronic parameterization  which consists of a
narrow--width vector state  plus the corresponding pQCD continuum does
remarkably well in reproducing the experimental shape of the Adler function.
This is very encouraging in view of the LMD approximation to \QCD which we
have suggested in the Introduction, namely to consider that the
large--$N_c$ spectrum consists of a prominent low--energy narrow state with
the rest of the narrow states lumped together as a
\QCD perturbative continuum.  It is this LMD approximation to
\QCD which we are now going to further explore in the following sections.

\end{itemize}

\section{Effective Lagrangians and Matching Conditions}
\setcounter{equation}{0}
\label{sec:three}

\noi 
Internal quark loops in Green's functions of colour--singlet quark
bilinear operators are suppressed in the large--$N_c$ limit. Quark degrees of
freedom appear only in the external loop. At low-energies, the external
momentum flowing into the quark loop is small. If one  ``assumes'' that
the dynamics of the gluons has a characteristic non--perturbative
scale
$\Lac\gtrsim 1\,\GeV$~\footnote{~Glueballs, for instance, are expected to
have masses
$\gtrsim 1$ GeV.}
 , it is natural to think of this quark loop as a
low--energy insertion of effective higher--dimensional composite quark
operators. These operators will be local at low--energy scales  below
$\Lac$. Furthermore, since the physics encoded in these effective
operators originates from integrating out degrees of freedom from
$\Lac$ to infinity, the QCD $U(3)_L\times U(3)_R$ flavor symmetry must
also be explicitly realized in the effective composite quark operators.
Na{\"\i}vely, one expects $\Lac^2$ and the 
$s_0$ of subsec.~\ref{subsec:twothree} which fixes phenomenologically the
onset of the pQCD continuum, to be rather similar.

In perturbation theory one can explicitly see the emergence of these
higher dimensional composite operators at very large orders in the QCD
coupling constant~\cite{VZ94,VZ96,PdeR97}. Although this is not a proof
of the {\it existence} of these operators, it is nevertheless reassuring to
know that there are physical systems which are well described by an
analogous mechanism. Indeed, a superconductor
\underline{is} an example of a physical system where the appearance of
higher--dimensional fermionic operators in the non--perturbative dynamics is
also hinted at by the very large orders of perturbation 
theory, (see e.g. ref.~\cite{Pe98}.) It is this kind of reasoning that leads
one to consider as a plausible starting {\it ansatz} to formulate the LMD
approximation to
\QCD, an effective Lagrangian of the form
\bea\label{eq:njl}
\cL_{\ENJL} & = &  \bar{q}(x)\Big[i\gamma^{\mu} D_{\mu} + i
(s-i\gamma_{5}\,p)\Big]q(x) \nn \\ & & +\frac{8\pi^2 G_S}{N_c
\Lambda_{\chi}^2}\
\sum_{\mbox{\rm \scriptsize a,b=flavour}}
\left(\bar q^a_R q^b_L \right)\
\left(\bar q^b_L q^a_R \right) \nn \\ & & - \frac{8\pi^2 G_V}{N_c
\Lambda_{\chi}^2}
\sum_{\mbox{\rm \scriptsize a,b=flavour}}
\left[ (\bar q^a_L \gamma^\m q^b_L)\ (\bar q^b_L \gamma_\m q^a_L) +
(\mbox{\rm L} \rightarrow \mbox{\rm R}) \right] +\cdots \,,
\eea   where $D_{\mu}= \partial_{\mu}- i l_{\mu} \frac{1-\gamma_5}{2}-i
r_{\mu}\frac{1+\gamma_5}{2}$ ;
$l_{\mu}=v_{\mu}-a_{\mu}$, $r_{\mu}=v_{\mu}+a_{\mu}$, $s$ and $p$ are
external matrix source fields and
$G_S$ and
$G_V$ are dimensionless couplings of $\cO(N_{c}^{0})$. Summation over
colour indices within brackets is understood. This effective Lagrangian
is a Nambu--Jona-Lasinio type Lagrangian~\cite{NJL61} in its extended
version~\cite{DSW85,BBdeR93,BdeRZ94,Bi96}. For the sake of brevity we
shall refer henceforth to the set of operators explicitly shown in eq.
(\ref{eq:njl}) as ENJL. The presence of the two types of four--quark
interactions is the minimum required to have non--trivial dynamics in
channels with the $J^P$ quantum numbers $0^-\,, 1^-\,,  0^+\,,
\annd\;1^+\,.$   Taken as it stands this Lagrangian, at leading order in
$1/\N$ and provided that the coupling $G_{S}$ becomes critical ($G_S >
1$), yields the wanted pattern of spontaneous chiral symmetry
breaking~\footnote{~The wanted pattern is the one which produces a sizable
quark condensate. Other possible realizations of spontaneous chiral
symmetry in QCD have been advocated by J.~Stern and
collaborators~\cite{FSS91,SSF93,KMSF95} and references therein. See
however~\cite{KdeR97}.}  and reproduces surprisingly well many of the low
energy properties of hadron phenomenology~\cite{BBdeR93,BdeRZ94,Bi96}  with
only three independent parameters: $G_S, G_V$ and
$\Lac$.

The ENJL Lagrangian, however, has basically two very important
drawbacks. On the one hand, although at $Q^2=0$  it produces the correct nonet
of Goldstone poles in the S--matrix that are color singlets and have the
right flavor quantum numbers; at higher energies it has {\it unconfined
quarks}, whose most evident effect is to create unphysical imaginary parts
in Green's functions. Furthermore, the {\it unconfined
quark--antiquark pairs} produce S--matrix elements which do not obey the
mesonic large--$N_c$ counting rules. On the other hand, there is no
explanation as yet on how one is supposed to do the {\it matching} of this
effective Lagrangian with short--distance QCD. As we shall see, the two
drawbacks are very much related to each other and to the {\it ellipsis} in
eq.~(\ref{eq:njl}).

\vspace*{0.5cm}

\subsection{The Adler Function of the ENJL--Lagrangian.}
\label{subsec:threeone}

\noi The explicit form of the Adler function predicted by the ENJL
Lagrangian has been calculated in ref.~\cite{BdeRZ94}. It reads
\be\label{eq:pienjl}
\cA_{\ENJL}(Q^2) = -Q^2\frac{d}{dQ^2}\Pi_{\ENJL}(Q^2)= -Q^2\frac{d}{dQ^2}\
\left( \frac{\Pivbar(Q^2)}{1 + Q^2 \frac{8 \pi^2 G_V}{\N \Lac^2}
  \Pivbar(Q^2)}\right) \quad ,
\ee  where $\Pivbar(Q^2)$ is given by a loop of constituent free quarks.
Since this function is ultraviolet divergent it requires the introduction
of a cut--off. We shall later come back to questions of ambiguities from the
choice of the cut--off; for the time being, we shall adopt the same proper
time regulator as in ref.~\cite{BdeRZ94}; then $\Pivbar(Q^2)$ has the form 
\be\label{eq:propertime}
\Pivbar(Q^2)=\frac{\N}{16\pi^2} \ 8\ \int_0^1 dx x (1-x) \Gamma(0,x_Q)\,,
\ee
 where $x_Q\equiv \frac{M_Q^2+Q^2x(1-x)}{\Lac^2}$ and
$\Gamma(n,\epsilon)\equiv
\int_{\epsilon}^{\infty} \frac{dz}{z} e^{-z}z^n$ is the incomplete Gamma
function. The constituent quark mass $M_Q$ is the non--trivial solution
($M_{Q}\ne 0$) to the gap equation
\be\label{eq:1stgapeq}
\frac{M_Q}{G_{S}}=M_{Q}\left[\exp{
\left(-\frac{M_{Q}^2}{\Lambda_{\chi}^2}\right)}-
\frac{M_{Q}^2}{\Lambda_{\chi}^2}
\Gamma(0,\frac{M_{Q}^2}{\Lambda_{\chi}^2})\right]\,.
\ee

The shape of the Adler function predicted by the ENJL Lagrangian is shown
in Fig.~5 (the dotted curve). As the figure clearly shows, ENJL does not
interpolate correctly the intermediate region between the very long--distance
regime and the short--distance regime; but what is more striking is that it
also fails to describe hadronic data already at such low energies as
$Q^2\lesssim 4 M_Q^2$ ($M_Q\simeq 300\,\MeV$). The na{\"\i}ve expectation that
this effective Lagrangian should be appropriate in describing the euclidean
behaviour of Green's functions up to
$Q^2\sim \Lac^2$, (since this is after all the scale that appears in the
denominators of the composite operators in eq.~(\ref{eq:njl}),) has
not been fullfilled~\footnote{~This solves the "paradox" raised in
ref.~\cite{YZ94}. There is no overlap between the region where the
ENJL--Lagrangian is operational and the region covered by the OPE.}. This
failure can be delayed to slightly higher
$Q$--values by introducing extra four--quark operators with higher
derivatives in the lines proposed in refs.~\cite{PP93,PP95}, but this
introduces more unknown couplings and does not address the lack of
confinement. 

\vspace{0.3cm}
\centerline{\epsfbox{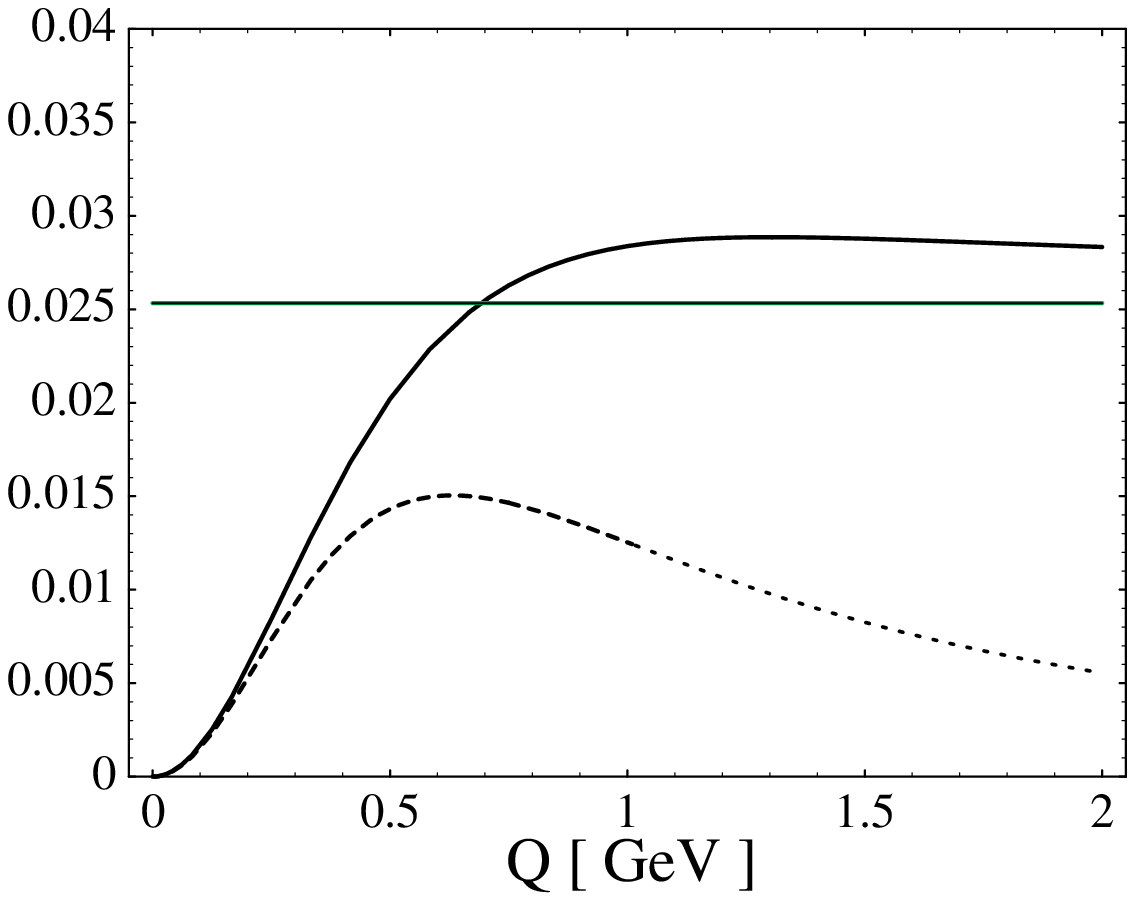}}
\vspace{0.2cm}

{\bf{Fig.~5}}
{\it{The Adler function predicted by the ENJL--Lagrangian in
eq.~(\ref{eq:njl}) (dashed line) compared with the Adler function from the
Aleph data (full line).}}

\vspace{0.3cm}

\noindent
What is the reason for this failure? In order to find guidance in
answering this question, let us compare $\cA_{\ENJL}(Q^2)$ in
eq.~(\ref{eq:pienjl}) with the expression which results from the LMD
approximation to
\QCD which we have been advocating in the previous sections, i.e. the
approximation where
\be\label{eq:largeN}
\frac{1}{\pi}\Imm\Pi_{\infty}(t)=2f_V^2 M_V^2 \delta(t-M_V^2) +
\frac{\N}{16\pi^2}\frac{4}{3} \theta (t-s_0)\,,
\ee 
and where for simplicity we are switching off $\als$ corrections in the
pQCD continuum term.  This spectral function produces in the euclidean
region the Adler function
\be\label{eq:a}
\cA_{\infty}(Q^2)=-Q^2\frac{d}{dQ^2}\left(\frac{2f_V^2 M_V^2}{M_V^2 +
Q^2}\right) + \frac{\N}{16\pi^2}\frac{4}{3} \frac{Q^2}{s_0 + Q^2}\,.
\ee  Clearly eq.~(\ref{eq:a}) does not look like eq.~(\ref{eq:pienjl}).
Even in the region of small $Q^2$ we see that, firstly,
eq.~(\ref{eq:pienjl}) lacks $Q^2/s_0$ contributions~\footnote{We would like
to issue a warning message here: the continuum contributes to some of the
$\cO (p^6)$ (and higher) couplings of the chiral expansion. Consequently,
saturating these coupling constants with only the resonance contribution, as
is sometimes done by invoking e.g. VMD arguments, may be misleading.} that are
present in eq.~(\ref{eq:a}); and, secondly, also the
$Q^2$ dependence of the resonance contribution is misrepresented in
eq.~(\ref{eq:pienjl}), at least insofar as
$\Pivbar(Q^2)$ is $Q^2$ dependent. All in all, the low--$Q^2$ dependence
of eq.~(\ref{eq:pienjl}) appears to be ``incorrect'' if expected to 
resemble the one in eq.~(\ref{eq:a}) but, as we shall see, in a way that can
be amended. The amendment will consist in adding the proper higher
dimensional composite operators. We shall see that this addition is nothing
but the implementation of the {\it matching conditions} inherent to any
effective Lagrangian construction.

\vspace*{0.5cm}

\subsection{Higher Dimension Operators and Confinement}
\label{subsec:threetwo}

\noi 
There will be two distinct classes of higher dimensional operators
according to whether they originate from the {\it continuum} or from the
prominent low energy narrow state, which we shall call for short the {\it
  resonance}. This is because chiral symmetry is realized differently in
these two regions of the spectrum. In the case of the {\it continuum},
chiral symmetry is manifest by construction and the new
higher--dimensional operators to be added to the Lagrangian in
eq.~(\ref{eq:njl}) will contain  solely combinations of external fields
suppressed by the corresponding powers of $s_0$, with chiral symmetry
realized in a linear way. A possible example is the operator
\be\label{op:cont}
\frac{1}{s_{0}}\tr\left\{F_{L\m\n}(x)\Box F_{L}^{\m\n}(x) +  F_{R\m\n}(x)
\Box F_{R}^{\m\n}(x)\right\}\,.
\ee
On the other hand, operators originating in the contribution from the lowest
{\it resonance} must have spontaneous chiral symmetry breaking built in,
and therefore they will appear as chirally symmetric in a nonlinear way;
i.e. they will eventually generate effective couplings  involving the
Nambu--Goldstone bosons encoded in the
$U$--matrix and suppressed by a characteristic mass scale proportional to  the
$Q\bar{Q}$ production threshold
$2 M_{Q}$. A possible example of an operator of this type is the dimension
six operator
\be\label{op:contnl}
\frac{1}{4M_{Q}^2}\tr\left\{F_{L\m\n}(x)U(x)^{\dagger}[\Box
F_{R}^{\m\n}(x)]U(x) +  F_{R\m\n}(x)U(x)[\Box
F_{L}^{\m\n}(x)]U(x)^{\dagger}\right\}\,.
\ee

In order to proceed to the explicit construction of these higher dimension
operators it helps to have yet another perspective on the failure of
eq.~(\ref{eq:njl}) to reproduce the expected structure of the LMD
approximation to
\QCD.
When computing the
imaginary part of $\Pi_{\ENJL}$ in eq.~(\ref{eq:pienjl}) one finds:
\be\label{eq:impi}
\frac{1}{\pi}\Imm\Pi(t)_{\ENJL}=
\frac{\frac{1}{\pi}{\rm Im}\Pivbar(t)} {\left(1-t \frac{8\pi^2G_V}{\N
\Lac^2}{\rm Re}\Pivbar(t)\right)^2 +
\left(t \frac{8\pi^2G_V }{\N \Lac^2} {\rm Im} \Pivbar(t)\right)^2}\,.
\ee The ENJL spectral function has a continuous shape, as shown in
Fig.~6, whereas the large--$N_c$ result should be a sum of narrow states;
or at least a dominant narrow state as in eq.~(\ref{eq:largeN}).

\vspace{0.3cm}
\centerline{\epsfbox{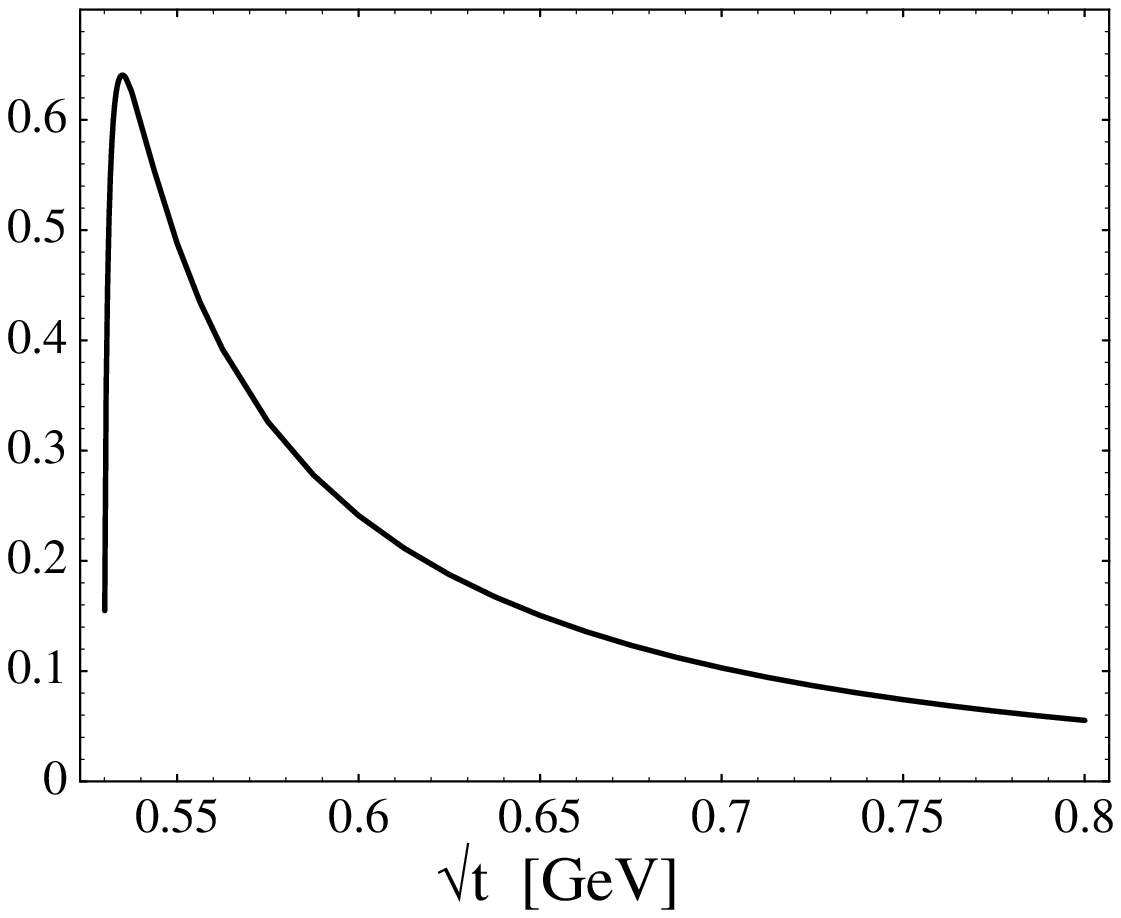}}
\vspace{0.2cm}

{\bf{Fig.~6}}
{\it{The vector spectral function predicted by the
ENJL--Lagrangian}~\cite{BdeRZ94}}

\vspace{0.3cm}

\noindent
The shape in Fig.~6 seems reminiscent of the one corresponding to a
``$\rho$--resonance'' with a ``finite width'' proportional to
\be
\gamma=t\frac{8\pi^2 G_{V}}{N_{c}\Lac^2}\Imm\bar{\Pi}(t)\,,
\ee  and, at first sight, it could even be taken as a success of ENJL. This
shape however is more of a {\it failure} than a success because it involves a
``finite width'' which is proportional to Im$\Pivbar(t)$ and therefore of
$\cO(N_{c}^0)$ while in the large--$N_c$ limit of QCD, which after all
$\cL_{\ENJL}$ in eq.~(\ref{eq:njl}) is supposed to approximate in some
way, it should be of $\cO(1/N_c)$. The wrong large--$\N$ behaviour of the
ENJL width is actually due to the fact that the ENJL spectral function has
on--shell
$Q \bar Q$ pairs (which involves an extra factor of $\N$), i.e. precisely
to the lack of {\it confinement}. What is needed here is to
find the extra physics which generates the formal limit $\Imm\Pivbar(t)
\to 0$ in a consistent fashion, for then as
$\gamma \to 0$
\be
\lim_{\gamma\ra 0}\frac{\gamma}{\left(1-t \frac{8\pi^2G_V}{\N
\Lac^2}{\rm Re}\Pivbar(t)\right)^2 + \gamma^2 } \ra
\pi\delta \left(1-t \frac{8\pi^2G_V}{\N
\Lac^2}{\rm Re}\Pivbar(t)\right)\,,
\ee 
and the narrow width limit would follow.

The obvious way to perform this limit that comes to one's mind is to take
the limit
$M_Q\to\infty$. This is actually what happens in \QCD in two
dimensions~\cite{'tH74b,CCG76}, where the mass of the quark is pushed to
infinity by the need to regulate the theory in the infrared region; and
in two dimensions this is in fact how confinement takes place. Here this
does not work because the constituent quark mass
$M_{Q}$ given by the dynamics of $\cL_{\ENJL}$ in eq.~(\ref{eq:njl}) is
finite. One can nevertheless proceed as follows. Let us consider  the
very low--$Q^2$ region. Since at
$Q^2=0$ the operators of eq.~(\ref{eq:njl}) produce the right physics,
one can start at $Q^2=0$ and work the way up by examining what
happens when trying an expansion in powers of
$Q^2/4M_Q^2$. Because  $M_Q$ is finite, the $Q^2$ dependence
of $\Pivbar(Q^2)$ is a consequence of the fact that
$\Imm\Pivbar(t)$ is nonvanishing. The $Q^2$ dependence of $\Pivbar(Q^2)$
originates from the production of constituent $Q\bar{Q}$--pairs in the
physical spectral function; in other words from the {\it lack of confinement}.
Indeed, the
$Q^2$ dependence is given by the dispersive integral
\be\label{eq:drbar}
\Pivbar(Q^2) = \Pivbar(0) - Q^2 \int_{4 M_Q^2}^{\infty} \frac{dt}{t
(t+Q^2)}
\frac{1}{\pi}\Imm\Pivbar(t)\,,
\ee  
where $\frac{1}{\pi}\Imm\Pivbar(t)$ is the spectral function associated with
the constituent $Q\bar{Q}$--pairs. It is in fact this dispersion relation that
holds the key to the remedy we are looking for. The crucial point is that
eq.~(\ref{eq:drbar}) shows that
$\Pivbar(Q^2)$ is analytic for $Q^2<4M_Q^2$. It can be expanded in powers
of
$Q^2/4M_Q^2$ and therefore all the ``wrong'' $Q^2$ dependence causing
trouble can be systematically removed by adding to the Lagrangian in
eq.~(\ref{eq:njl}) an {\it infinite} set of local operators involving
external fields of dimension $n\ge 6$:
\be\label{eq:external}
\cL_{\res} = \sum_{n=6}^{\infty} c_n \cO_n^{\res}.
\ee 
The lowest dimension is six because the first operator
corresponds to the first derivative of the self--energy function
$\bar{\Pi}(Q^2)$. The coefficients
$c_n$ in front of these local operators are known since they are adjusted
so that all powers of
$Q^2/4M_Q^2$ coming from the Taylor expansion of $\Pivbar(Q^2)$ cancel
out in their contribution to the Adler function in eq.~(\ref{eq:pienjl}).
For instance an explicit  operator of dimension six which
removes the ``wrong'' non--confining dependence of $\cO(Q^2)$ is
\be
\label{eq:externalone}
\cO_{6} = \bar \Pi'(0)\
\tr\left\{U^{\dagger} \left(D^\al F_{R}^{\m\n}\right)U \left(D_\al F_{L\m\n}
\right)\right\}\,,
\ee  where as usual the trace is in flavour space, and $\bar \Pi'(0)$ stands
for
\be\label{eq:coupling6}
\bar \Pi'(0)\equiv \frac{d\bar\Pi}{dQ^2}(0)=\frac{-N_c}{16\pi^2 M_{Q}^{2}}
\frac{4}{15}\Gamma(1,\epsilon)\,.
\ee
The presence of $\Gamma(1,\epsilon)$ in
eq.~(\ref{eq:coupling6}) is due to the proper time regularization which
we have adopted to define $\bar{\Pi}(Q^2)$ in eq.~(\ref{eq:propertime}).
The ``wrong'' non--confining $Q^2$ dependence is regularization
dependent and therefore, the counterterms which are added to cancel them
will be regularization dependent as well. The net physical effect when the
{\it infinite} set of local
operators in eq.~(\ref{eq:external}) is added, is formally the same as
letting
$M_{Q}\ra\infty$ in the $\frac{1}{\pi}\Imm\bar{\Pi}(t)$ spectral
function, namely
$\Pivbar(Q^2)\ra\Pivbar(0)$.

\vspace*{0.5cm}

\subsection{Higher Dimension Operators and Matching Conditions}
\label{subsec:threethree}

\noi
The higher dimension operators in eq.~(\ref{eq:external}) can also
be viewed as part of the {\it matching} conditions which are required on the
ENJL Lagrangian in eq.~(\ref{eq:njl}) to become an effective Lagrangian
describing a narrow width confined vector state. Indeed, once these
operators are included, effectively one is doing the following
replacement in eq.~(\ref{eq:pienjl}):
\be
\label{eq:piconfinedexpanded}
\Pi_{\ENJL}(Q^2) \ra \Pivbar(0) \sum_{n=0}^{\infty}
\left(- Q^2 \frac{8 \pi^2 G_V}{\N \Lac^2} \Pivbar(0)\right)^n\,,
\ee  which is nothing but the expansion around $Q^2=0$ of
\be
\label{eq:piconfined}
\Pi_{\eff}(Q^2) = \frac{\Pivbar(0)}{1 + Q^2 \frac{8 \pi^2 G_V}{\N
\Lac^2}\Pivbar(0)}\,,
\ee
and  has now the form produced by an infinitely narrow state, i.e.
\be\label{narrow}
\Pi_{\eff}(Q^2)=\int_0^{\infty} \frac{dt}{t + Q^2} \frac{1}{\pi} {\rm
Im}\Pi_{\eff}(t) \ ,
\ee
 with
\be
\frac{1}{\pi}\Imm\Pi_{\eff}(t) = \frac{\N \Lac^2 }{8
\pi^2 G_V}
\delta(t-\N \Lac^2 (8 \pi^2 G_V
  \Pivbar(0))^{-1})\,.
\ee
This is precisely the desired result. It leads to a hadronic contribution
to the Adler function
\be
\cA_{\eff}(Q^2)=-Q^2\frac{d}{dQ^2}\left(\frac{2f_V^2 M_V^2}{M_V^2 +
Q^2}\right)\,,
\ee
with
\be
2f_{V}^2M_{V}^2=\frac{\N \Lac^2 }{8\pi^2 G_V}\quad\annd\quad
M_V^2= \N \Lac^2 (8 \pi^2 G_V
  \Pivbar(0))^{-1}\,;
\ee
i.e., the same contribution as the LMD approximation to \QCD formulated in
eq.~(\ref{eq:a}).  

At this point the reader may perhaps wonder about what all this
improvement on ENJL has really brought us since it looks as if all we have
accomplished is to reproduce eqs.~(\ref{eq:largeN}) and (\ref{eq:a}),
which after all we already knew from large--$\N$ QCD arguments in the
first place. The point, as we shall further develop in the next sections,
is that imposing that the low--energy physics be governed by the
effective Lagrangian in eq.~(\ref{eq:njl}) with the {\it infinite} set of
{\it matching} operators of higher dimension understood, constrains
the parameters of the dominant low--energy states, which would otherwise be
arbitrary. It relates, in particular, the lowest vector
state to the lowest axial--vector state in a way which we will
describe next.

\section{The Axial--Vector Channel}
\setcounter{equation}{0}
\label{sec:four}

\noi   We shall now be concerned with two-point functions
\be\label{eq:tpfA}
\Pi_{\mu\nu}^{A}(q)_{ab}=i\int d^{4}xe^{iq\cdot x}<0|T\{A_\mu^{a}(x)
A_\nu^{b}(0)\}|0>
\ee
of axial--vector quark currents
\be\label{eq:Acur}
A_\mu^{a}(x)=\bar q (x)\gamma_{\mu}\gamma_{5}
\frac{\lambda^{a}}{\sqrt{2}}q(x)\,,
\ee
where $\lambda^{a}$, as before in eq.~(\ref{eq:Vcur}), are Gell-Mann
matrices acting on the flavour triplet of $u$, $d$, $s$ light quarks. In
the chiral limit where the light quark masses are set to zero, these
two--point functions depend only on one invariant function ($Q^2 =-q^2\ge
0$ for
$q^2$ spacelike)
\be\label{eq:itpfA}
\Pi_{\mu\nu}^{A}(q)_{ab}=(q_\mu q_\nu -g_{\mu\nu}q^2)
\Pi_{A}(Q^2)\delta_{ab}\,.
\ee The axial--vector two--point functions predicted by the ENJL Lagrangian
have been discussed in ref.~\cite{BdeRZ94}. The self--energy
function $\Pi_{A}(Q^2)$ has the following form
\be\label{eq:axialpi}
\Pia(Q^2) = \frac{\bar \Pi_A(Q^2)}{1 + Q^2 \frac{8 \pi^2
    G_V}{\N \Lac^2}\ \bar \Pi_A(Q^2)}\,,
\ee  where $\bar \Pi_A(Q^2)$ is now given by
\be\label{eq:axialbar}
\bar \Pi_A(Q^2) = \bar{\Pi}(Q^2) + \frac{f^2(Q^2)}{Q^2}\,,
\ee
with $\bar{\Pi}(Q^2)$ the same function as the one defined in
eq.~(\ref{eq:propertime}) and
\be  f^2(Q^2)=\frac{\N}{16\pi^2}\ 8 M_Q^2\ \int_0^1 dx \Gamma(0,x_Q)\,.
\ee
It is convenient to introduce the function
\be
g_{A}(Q^2)=\frac{1}{1+(G_{V}/\Lac^2)4M_{Q}^{2}\int_{0}^{1}
dx\Gamma(0,x_{Q})}\,,
\ee
and then rewrite $\Pi_{A}(Q^2)$ as follows
\be \label{eq:axialnaive}
\Pi_A(Q^2) = \frac{\Pivbar(Q^2) g_{A}^2(Q^2)}{1+ Q^2 \frac{8\pi^2G_V}{\N
\Lac^2}
  \Pivbar(Q^2) g_A(Q^2)}\ +\ \frac{f^2(Q^2) g_A(Q^2)}{Q^2}\,.
\ee As in the case of the vector channel one notices that
eq.~(\ref{eq:axialnaive}) could match onto an axial Adler function
\be
\label{eq:adleraxiallargeN}
\cA_A(Q^2)= - Q^2\frac{d}{dQ^2}\left( \frac{2f^2_{\pi}}{Q^2} +
\frac{2f_A^2
    M_A^2}{M_A^2+Q^2}\right) + \frac{\N}{16 \pi^2} \frac{4}{3}
    \frac{Q^2}{s_0+Q^2}\,,
\ee  
generated by a narrow width spectral function corresponding to the LMD
approximation in
\QCD i.e.,
\be\label{eq:axiallargeN}
\frac{1}{\pi}\Imm\Pi_A(t) = 2f_{\pi}^2 \delta(t) + 2f_A^2 M_A^2
\delta(t-M_A^2) + \frac{\N}{16\pi^2} \frac{4}{3} \theta(t-s_0)\,,
\ee  provided that both $\Pivbar(Q^2)$ and $f^2(Q^2)$ get their $Q^2$
dependence, induced by $Q\bar{Q}$ discontinuities, canceled by the inclusion
of appropriate higher dimensional composite operators in the external fields.

We emphasize that the procedure of introducing an infinite number of
local operators adds no unknown parameters, as the coefficients
accompanying the operators are fixed by the coefficients of the Taylor
expansion at zero momentum of the functions
$\Pivbar(Q^2)$ and $f^{2}(Q^2)$ which are explicitly known. Furthermore,
one also has to add the operators that fix the contribution of the
continuum, as was explained in the case of the vector channel in
subsec.~\ref{subsec:threetwo}. Once this is implemented the masses and
coupling constants of the dominant low--energy states in the vector and
axial--vector channels are completely fixed in terms of the parameters
$G_S,G_V$ and $\Lac$ in a way which we summarize below.

\begin{description}

\item{i) {\sc The gap equation.}}\\ The gap equation in
eq.~(\ref{eq:1stgapeq}) relates $G_S$ to the ratio of the constituent quark
mass $M_Q$ and the scale $\Lac$
\be\label{eq:gapeq}
\frac{1}{G_S}=\epsilon\Gamma(-1,\epsilon)\,,\qquad\with\qquad
\epsilon\equiv
\frac{M_{Q}^{2}}{\Lac^2}\,.
\ee

\item{ii) {\sc The coupling constant $g_{A}$.}}\\
The coupling constant $g_A$ is governed by the size of the $G_V$ coupling
\be g_A\equiv g_A(0)=\left( 1 + 4 G_{V}\epsilon
  \Gamma(0,\epsilon)\right)^{-1}\,.
\ee

\item{iii) {\sc Masses.}}\\ The masses of the vector and axial--vector
states are then fixed as follows
\be\label{eq:MVMA}
 M_V^2=\frac{3}{2}
\frac{\Lac^2}{G_V} \frac{1}{\Gamma(0,\epsilon)} = 6 M_Q^2
  \frac{g_A}{1-g_A}\,,\quad
\annd\quad
M_{A}^{2}= \frac{M_{V}^{2}}{g_A}\,.
\ee

\item{iv) {\sc The Coupling Constants.}}\\ The coupling constants of the
lowest narrow states are then also fixed:
\bea  \label{eq:fpi}
f_{\pi}^2 & = &
\frac{N_c}{16\pi^2}4M_{Q}^2g_{A}\Gamma(0,\epsilon)\,,
\\  \label{eq:fV}
f_{V}^2 & = &
\frac{N_c}{16\pi^2}\frac{2}{3}\Gamma(0,\epsilon)\,,\\ \label{eq:fA}
f_{A}^2 & =
& g_{A}^2 f_{V}^2\,.
\eea

\end{description}

\noi With these results, the 1st and 2nd Weinberg sum rules, which in the
case of LMD imply the well known relations
\bea
\label{eq:1stwsr}  f_{\pi}^2 +f_{A}^{2}M_{A}^{2} & = &
f_{V}^{2}M_{V}^{2}\,,
 \\
\label{eq:2ndwsr} f_{A}^{2}M_{A}^{4} & = & f_{V}^{2}M_{V}^{4}\,,
\eea are then \underline{automatically} satisfied. It appears then, that just
removing the {\it non--confining} terms in the vector and axial--vector
two--point functions predicted by the usual ENJL--Lagrangian guarantees their
correct {\it matching} to the QCD short--distance behaviour. This shows,
in retrospect,  that the choice of an initial ENJL--{\it ansatz} as a
starting effective Lagrangian was quite a good one. However, as we shall
later see, the removal of the {\it non--confining} contributions
in Green's functions beyond two--point functions is not enough to
guarantee, in general, the correct {\it matching} to the QCD short--distance
behaviour and further local operators have to be introduced.

\vspace*{0.5cm}

\subsection{The Electromagnetic $\pi^{+}-\pi^{0}$ Mass--Difference}
\label{subsec:fourone}

\noi In the chiral limit, and in the presence of electromagnetic
interactions to lowest order in the fine structure constant
$\alpha$, the  hadronic spectral functions in eqs.~(\ref{eq:largeN})  and
(\ref{eq:axiallargeN}) induce an electromagnetic mass for the charged 
pion, which is finite when the spectral functions obey the 1st and 2nd
Weinberg sum rules~\footnote{~For an introduction to this subject, see
e.g. the lectures in ref.~\cite{deR98}, where references to the earlier
work are also given.}. This results in the expression
\be  m_{\pi^+}^{2}\vert_{\EM}=
\frac{\alpha}{\pi}\frac{3}{4}\frac{M_{A}^2 M_{V}^2}{M_{A}^2 -M_{V}^2}
\log\frac{M_{A}^2}{M_{V}^2}\,,
\ee which in terms of the parameters $M_{Q}$ and $g_{A}$ defined above
reads
\be\label{eq:piem}
m_{\pi^{+}}^{2}\vert_{\EM}=\frac{\alpha}{\pi}\frac{9}{2}M_{Q}^2
\frac{g_{A}}{(1-g_{A})^2}\log\frac{1}{g_{A}}\,.
\ee
We show below in Fig.~7 a two--dimensional plot of
the resulting electromagnetic pion mass difference $\Delta m_{\pi}\equiv
m_{\pi^+}-m_{\pi^0}$ as a function of $M_Q$ and
$g_A$ when these parameters are varied in the ranges $200\,\MeV\le M_Q\le
400\,\MeV$ and
$0.4\le g_{A}\le 0.6$. Experimentally~\cite{PDB}, $\Delta m_{\pi}=
(4.5936\pm 0.0005)\,\MeV$. This corresponds to the {\it plateau} shown in
Fig.~7. The intersection of eq.~(\ref{eq:piem}) with this {\it plateau}
results in a line which correlates $M_Q$ and $g_{A}$. The figure shows that
a very reasonable choice of values for these two parameters can reproduce
the experimental result. We shall be more precise about the values of
these parameters in subsection 5.2     

\vspace{0.3cm}
\centerline{\epsfbox{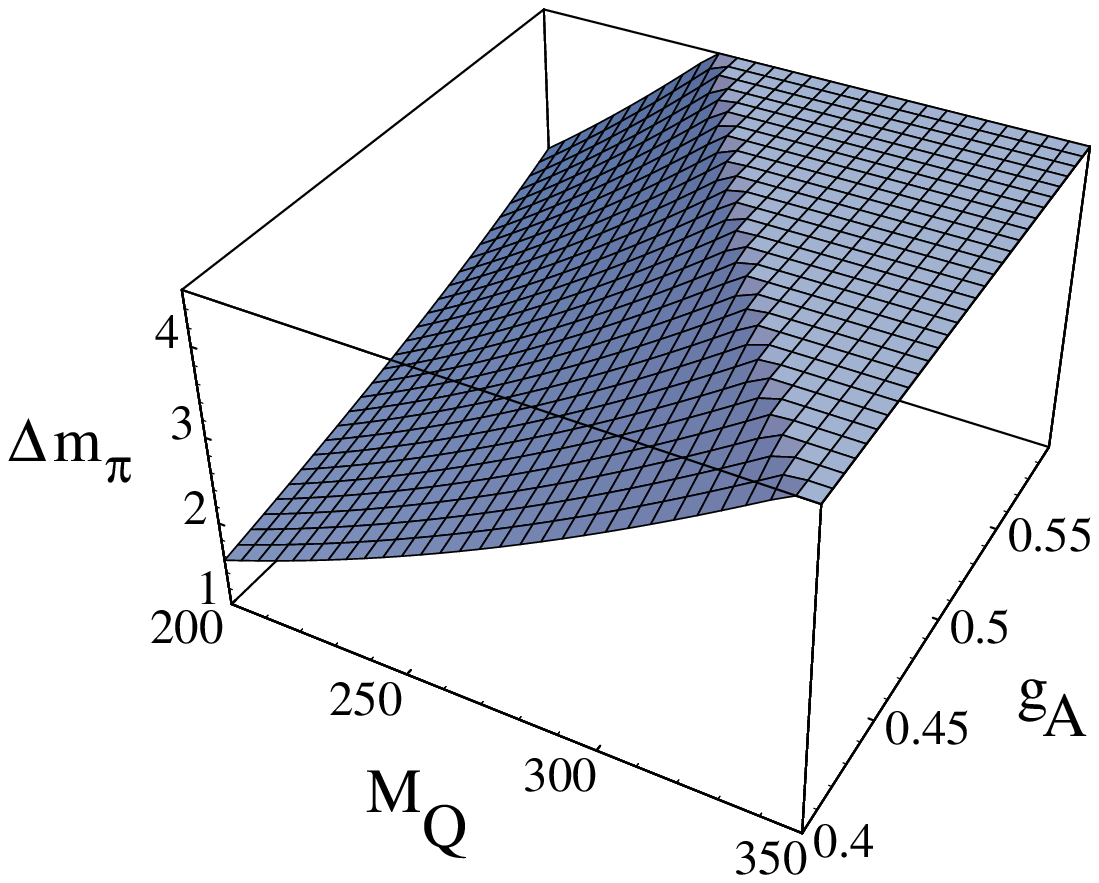}}
\vspace{0.2cm}

{\bf{Fig.~7}}
{\it{The electromagnetic pion mass difference $\Delta m_{\pi}$ in $\MeV$
versus
$g_A$ and
$M_{Q}(\MeV)$ }}

\vspace{0.3cm}
\section{The Low--Energy Effective Lagrangian}
\setcounter{equation}{0}
\label{sec:five}

\noi The degrees of freedom of the resulting effective low--energy Lagrangian
are a pseudoscalar Goldstone field matrix, a vector field matrix
$V(x)$, a scalar field matrix
$S(x)$, and an axial--vector field matrix $A(x)$. They correspond, therefore,
to the wanted LMD spectrum of
$0^{-}$, $1^{-}$, $0^{+}$, and $1^{+}$ hadronic states. We postpone the
technical discussion concerning the removal of
$Q\bar{Q}$ discontinuities in Green's functions with external scalar sources
to another paper. Here, we shall limit ourselves to quote the results we
obtain for the effective $\cO(p^4)$ chiral Lagrangian of the Goldstone modes
after integrating out the physical scalar fields. It is this dynamical effect
which produces the
$L_{5}$ and
$L_{8}$ couplings as well as part of the $\tilde{L}_{3}$ coupling  which we
give below. The remaining interactions between the  pseudoscalar Goldstone
field matrix and the vector and axial--vector field matrices produce the rest
of the couplings in the following effective Lagrangian, valid up to
$\cO (p^4)$ and also for some of the higher order terms,
\bea\label{eq:Leff4}
\tilde{\cL} _{\eff} & = &  \frac{1}{4} f_{\pi}^2\, \left[ \tr
\left(D_{\mu}UD^{\mu}U^{\dagger}\right) + \tr \left( \chi
  U^{\dagger}+U^{\dagger} \chi \right)\right] +\mbox{\rm e}^2\,C\tr
(Q_{R}UQ_{L}U^{\dagger}) 
\nn \\ & & -\frac{1}{4}\tr\,[V_{\mu\nu}V^{\mu\nu}-2M_{V}^2 V_{\mu}V^{\mu}]
 \nn \\ & &  -\frac{1}{4}\tr\,[A_{\mu\nu}A^{\mu\nu}-2M_{A}^2 A_{\mu}A^{\mu}] 
\nn
\\ & &  -\frac{1}{2\sqrt{2}}\left[f_{V}\tr\left(
V_{\mu\nu}f_{+}^{\mu\nu}\right)+i\tilde{g}_{V}\tr\left( V_{\mu\nu}[\xi^{\mu},
\xi^{\nu}]\right)+f_{A}\tr\left( A_{\mu\nu}f_{-}^{\mu\nu}\right)\right] \nn
\\ & & 
   + \tilde{L}_1 \left(\tr\, D_{\mu}U^{\dagger} D^{\mu}U\right)^2+
2\tilde{L}_1
\tr\left( D_{\mu} U^{\dagger} D_{\nu}U\tr\, D^{\mu}U^{\dagger}D^{\nu}U\right)
\nn
\\ & &  +\tilde{L}_3
\tr\left( D_{\mu}U^{\dagger} D^{\mu}UD_{\nu}U^{\dagger} D^{\nu}U
\right) \nn \\ & & + L_5\tr\left[D_{\mu}U^{\dagger} D^{\mu}U
\left(\chi^{\dagger}U+U^{\dagger}\chi\right)\right] + 
L_8\tr\left(\chi^{\dagger}U \chi^{\dagger}U +\chi U^{\dagger}\chi
U^{\dagger}\right) \nn \\ & & 
 -i\tilde{L}_9 \tr\left(F^{\mu\nu}_R D_{\mu}U  D_{\nu}U^{\dagger}
+F^{\mu\nu}_L D_{\mu}U^{\dagger} D_{\nu}U\right)+L_{10} \tr\left( U^{\dagger}
F^{\mu\nu}_R U F_{L\mu\nu}\right) \nn \\ & & 
+H_{1}\tr\left(F_{R}^{\mu\nu}F_{R\mu\nu}+F_{L}^{\mu\nu}F_{L\mu\nu}\right)+
H_{2}\tr\left(\chi^{\dagger}\chi\right)\,.
\eea   Here,
\be V^{\mu\nu}= \nabla^{\mu}V^{\nu}-\nabla^{\nu}V^{\mu},\qquad A^{\mu\nu}=
\nabla^{\mu}A^{\nu}-\nabla^{\nu}A^{\mu}\,;
\ee with $\nabla$ the covariant derivative defined as
\be \nabla_{\mu}R\equiv \partial_{\mu}R+[\Gamma_{\mu},R]\,,
\ee  and $\Gamma_{\mu}$ the connection
\be
\Gamma_{\mu}=\frac{1}{2}\left\{\xi^{\dagger}\left[\partial_{\mu}
-i(v_{\mu}+a_{\mu})\right]\xi+
\xi\left[\partial_{\mu} -i(v_{\mu}-a_{\mu})\right]\xi^{\dagger}\right\}\,.
\ee  We recall that $v_{\mu}$ and $a_{\mu}$ are external vector and
axial--vector matrix field sources, (not to be confused with the hadronic
matrix fields $V_{\mu}$ and $A_{\mu}$,) and $F_{L}^{\mu\nu}$ and
$F_{R}^{\mu\nu}$ are the external field strength matrix tensors associated
with the
$l_{\mu}=v_{\mu}-a_{\mu}$ and $r_{\mu}=v_{\mu}+a_{\mu}$ sources. We are also
using the notation $U=\xi\xi$ with
\be   f_{\pm}^{\mu\nu}=\xi F_{L}^{\mu\nu}\xi^{\dagger}\pm \xi^{\dagger}
F_{R}^{\mu\nu}\xi\qquad\annd\qquad \xi_{\mu}=i\xi^{\dagger}D_{\mu}
U\xi^{\dagger}\,.
\ee
This is the effective Lagrangian which results from the ENJL--{\it ansatz}
after removal of {\it non--confining} $Q\bar{Q}$ discontinuities. As already
mentioned, this guarantees the correct {\it matching} to the QCD
short--distance behaviour for two--point functions only. The couplings
$\tilde{g}_{V}$, $\tilde{L}_{9}$, and $\tilde{L}_{1}$, $\tilde{L}_{3}$ are
correlated with three-- and four--point functions which still require the
introduction of extra higher dimension operators to guarantee the correct
{\it matching} to the leading QCD short--distance behaviour.
This is why we put a {\it tilde} on them, and we shall discuss how they have
to be modified in the next subsection. All the constants appearing in the
Lagrangian in eq.~(\ref{eq:Leff4})  are functions of
$G_{S}$, $G_{V}$, and $\Lambda_{\chi}$ alone or equivalently $g_{A}$,
$M_Q$ and $\Gamma_{0}\equiv\Gamma(0,\epsilon)$. The masses $M_{V}^2$ and
$M_{A}^2$ have already been given in eq.~(\ref{eq:MVMA}), and the couplings
$f_{\pi}^2$,
$f_{V}^2$ and $f_{A}^2$ in eqs.~(\ref{eq:fpi}), (\ref{eq:fV}) and
(\ref{eq:fA}). The constant $C$ is related to the electromagnetic $\pi^+
-\pi^0$ mass difference in eq.~(\ref{eq:piem}), 
\be
\frac{2e^2 C}{f_{\pi}^2}=m_{\pi^+}^2\vert_{\EM}\,.
\ee
One could in principle consider as well terms of $\cO(e^2 p^2)$ but this
goes beyond the scope of this paper.
       At this stage i.e., after removing
$Q\bar{Q}$ discontinuities in the initial ENJL--{\it ansatz}, the result for
the VPP coupling
$\tilde{g}_{V}$ is 
\be
\tilde{g}_{V}^2=
\frac{N_c}{16\pi^2}\frac{1}{6}(1-g_{A}^2)^2\Gamma_{0}\,,
\ee  and the  results for the other couplings in eq.(\ref{eq:Leff4}) are:
\bea\label{eq:L1}  \tilde{L}_{1} & = &
\frac{N_c}{16\pi^2}\frac{1}{48}(1-g_{A}^2)^2\Gamma_{0}\,, \\
\label{eq:L3} \tilde{L}_{3} & = & \frac{N_c}{16\pi^2}\frac{1}{8}
\left[-(1-g_{A}^2)^2 + 2g_{A}^4\right]\Gamma_{0}\,, \\ 
\label{eq:L5} L_{5} & = &
\frac{N_c}{16\pi^2}\frac{1}{4}g_{A}^3 \Gamma_{0}\,, \\ 
\label{eq:L8} L_{8} & = &
\frac{N_c}{16\pi^2}\frac{1}{16}g_{A}^2 \Gamma_{0}\,, \\ 
\label{eq:L9} \tilde{L}_{9} & = &
\frac{N_c}{16\pi^2}\frac{1}{6}(1-g_{A}^2)\Gamma_{0}\,, \\
\label{eq:L10} L_{10} & = &
\frac{N_c}{16\pi^2}\frac{-1}{6}(1-g_{A}^2)\Gamma_{0}\,.
\eea  

Integrating out the vector and axial--vector fields in the Lagrangian of
eq.~(\ref{eq:Leff4}) produces terms of
$\cO(p^6)$. However, to that order, there are two extra sources of other
possible contributions which must be taken into account. One is the effect of
integrating out the vector and axial--vector fields in the other
$\cO(p^3)$ interaction terms which are not included in 
$\cL _{\eff}$ in eq.~(\ref{eq:Leff4}), (terms like e.g.
$\epsilon_{\mu\nu\rho\sigma}\tr(V^{\mu}\xi^{\nu}\xi^{\rho}\xi^{\sigma})$;)
the other source, as already mentioned, is the effect of new
operators coming from the {\it matching} of Green's functions beyond
two--point functions to short--distance QCD.

The Adler two--point functions which we have discussed in the previous
sections, i.e. the expressions in eqs.~(\ref{eq:a}) and
(\ref{eq:adleraxiallargeN}) with the masses and couplings as given in
eqs.~(\ref{eq:MVMA}) and (\ref{eq:fpi}), (\ref{eq:fV}), (\ref{eq:fA}), are
the result of the full resummation of all the relevant terms in the effective
Lagrangian. Expanding these Adler functions in powers of $Q^2$ is, of course,
another way of obtaining the couplings of the corresponding operators in the
effective Lagrangian which govern the behaviour of these functions.

\vspace*{0.5cm}

\subsection{Dynamical Symmetries of the Low--Energy Effective Lagrangian}
\label{subsec:fiveone}

\noi  The results we find at this level for the low--energy couplings show
already a lot of interesting symmetries. It is worthwhile describing them and
comparing them with the dynamical relations discussed in ref.~\cite{EGLPR89}.
For the sake of comparison it is more convenient to rewrite the results we
have found for the couplings
$f_{V}$, $f_{A}$ and $\tilde{g}_{V}$ in the following way
\bea \label{eq:fVa}  f_{V} & = &
\sqrt{\frac{1}{1-g_A}}\frac{f_{\pi}}{M_{V}}\,,
\\ \label{eq:fAa} f_{A} & = & g_{A}
\sqrt{\frac{1}{1-g_A}}\frac{f_{\pi}}{M_{V}}\,,
\\ \label{eq:gVa} \tilde{g}_{V} & = &
\frac{1+g_A}{2}\sqrt{1-g_A}\frac{f_{\pi}}{M_{V}}\,.
\eea   The results for the couplings in eqs.~(\ref{eq:L1}) to (\ref{eq:L10})
can then be written as follows:
\bea
\label{eq:L1couplings} 
\tilde{L}_{1} & = & \frac{1}{8}\tilde{g}_{V}^2\,,
\\
\label{eq:L3couplings} \tilde{L}_{3} & = & -6\tilde{L}_{1}+g_{A}L_{5}\,, \\
L_{5} & = & 4g_{A}L_{8}\,, \\ L_{8} & = &
\frac{N_c}{16\pi^2}\frac{1}{16}g_{A}^2\Gamma_{0}\,, \\
\label{eq:L9couplings} \tilde{L}_{9} & = &
\frac{1}{2}f_{V}\tilde{g}_{V}\,,
\\
\label{eq:L10couplings} L_{10} & = & \frac{-1}{4}(f_{V}^2-f_{A}^2)\,.
\eea

\noi Some of the relations encoded in the Lagrangian (\ref{eq:Leff4}) turn
out to be common to those discussed in ref.~\cite{EGLPR89} in connection with
the formulation of chiral Lagrangians of vector and axial--vector mesons
obeying short--distance QCD constraints. They are the following:
\begin{itemize}

\item The 1st and 2nd Weinberg sum rules in eqs.(\ref{eq:1stwsr}) and
(\ref{eq:2ndwsr}).
 
\item The relationship between the constant $L_{10}$ and the  $f_{V}$ and
$f_{A}$ couplings in eq.(\ref{eq:L10couplings}).

\item The relationship between the constant $\tilde{L}_{9}$ and the $f_{V}$
and
$\tilde{g}_{V}$ couplings in eq.(\ref{eq:L9couplings}).

\item The relations:
\be\label{eq:froissart} 
\tilde{L}_{1}=\frac{1}{8}\tilde{g}_{V}^2\,,\qquad \annd \qquad
\tilde{L}_{3}\vert{\mbox{\tiny\rm vector}}=-6\tilde{L}_{1}\,,
\ee  where $\tilde{L}_{3}\vert{\mbox{\tiny\rm vector}}$ means the
contribution induced by vector exchange to $\tilde{L}_{3}$. In
ref.~\cite{EGLPR89} it is shown that these two relations follow from the
assumption that the forward scattering amplitudes of two pseudoscalar mesons
obey the Froissart bound~\cite{Fr61,Ma63}. 

\end{itemize}

\noi The fact that the effective Lagrangian in (\ref{eq:Leff4}) obeys
automatically these relations shows again that the ENJL--Lagrangian {\it
ansatz} plus the corresponding set of local operators required to remove the
non--confining
$Q\bar{Q}$--discontinuities which it generates is already a very good 
effective Lagrangian to describe the LMD approximation to  \QCD.   There
are however other relations in ref.~\cite{EGLPR89} which the effective
Lagrangian in (\ref{eq:Leff4}) does not quite satisfy. Let us describe them. 
 
\begin{description}

\item{i)} In ref.~\cite{EGLPR89} it is shown that the assumption that the
pion electromagnetic form factor
$F(q^2)$  with the charge normalization condition $F(0)=1$, obeys an
unsubtracted dispersion relation, and the saturation of the dispersive
integral with just the
$\rho$--pole leads to the relation 
\be\label{eq:emffndr}  f_{V}g_{V}\frac{M_{V}^{2}}{f_{\pi}^{2}}=1\,,
\ee with $g_{V}$ the physical $\rho\pi\pi$ coupling constant.

\item{ii)} A similar assumption of  an unsubtracted dispersion relation for
the axial form factor in $\pi\ra e\nu_{e}\gamma$, with the same saturation
assumption of the dispersive integral by a resonance pole was shown in
\cite{EGLPR89} to lead to the relation
\be\label{eq:affndr}  f_{V}=2g_{V}\,,
\ee  which is sometimes called the KSFR--relation~\cite{KS66,RF66}.

\end{description}

\noi  The combination of i) and ii) results in
\be
\label{eq:ksfr}  f_{V}=2g_{V}=\sqrt{2}\frac{f_{\pi}}{M_{V}}\,.
\ee The combination of i), ii) and the two Weinberg sum rules
 in eqs.(\ref{eq:1stwsr}) and (\ref{eq:2ndwsr}) imply
\be\label{eq:wam}   M_{A}=\sqrt{2}M_{V}
\ee   and therefore
\be\label{eq:low} 
\left(\pi_{+}^{2}-\pi_{0}^{2}\right)_{\EM}=
\frac{\alpha}{\pi}\frac{3}{2}M_{V}^{2}\log 2\,.
\ee  With $f_{\pi}$ and $M_{V}$ as independent variables, the other couplings
are then all fixed:
\be
\label{eq:couplingsEGLPR}  f_{V}=\sqrt{2}\frac{f_{\pi}}{M_{V}}\,, \quad
g_{V}=\frac{1}{\sqrt{2}}
\frac{f_{\pi}}{M_{V}}\,, \quad \annd\quad f_{A}=\frac{f_{\pi}}{M_{A}}\,.
\ee When compared to the results in eqs.~(\ref{eq:fVa}), (\ref{eq:gVa}) and
(\ref{eq:fAa}) it looks as if there is no unique solution for
$g_{A}$ which can reproduce all these constraints  simultaneously. The value
$g_{A}=\frac{1}{2}$ reproduces the result
$M_{A}=\sqrt{2}M_{V}$ and the couplings for $f_{V}$ and $f_{A}$ in
eq.~(\ref{eq:couplingsEGLPR}), but not $g_{V}$. On the other hand, the
electromagnetic pion form factor constraint in eq.~(\ref{eq:emffndr}) can
only be satisfied if $g_{A}=1$, while the axial--form factor constraint in
eq.~(\ref{eq:affndr}) requires
$g_{A}=0$; but both these two extreme values are excluded by general
large--$N_c$ arguments~\cite{KdeR97}. 

From the description above, there follows that the question of whether or not
the electromagnetic pion form factor and the axial form factor in
$\pi\ra e\nu_{e}\gamma$ obey unsubtracted dispersion relations is an
important issue. In QCD it is indeed expected that these form factors do obey
unsubtracted dispersion relations, and there is no reason that the
restriction of full QCD to \QCD with LMD spoils this
property~\footnote{~Short--distance QCD arguments for a $1/Q^2$ fall--off
of the electromagnetic pion form factor can be found in ref.~\cite{LS92}.
The $1/Q^2$ fall--off also follows rigorously from an analysis~\cite{MOU97} of
QCD short--distance constraints on three--point functions saturated by a
vector state and pseudoscalar Goldstones only.}. Let us then consider the
electromagnetic pion form factor
$F(Q^2)$ in somewhat more detail. The effective Lagrangian in
eq.~(\ref{eq:Leff4}) gives:
\be  \tilde{F}(Q^2)  =  1
 -2\tilde{L}_{9}\frac{M_{V}^{2}}{f_{\pi}^2}\frac{Q^2}{M_{V}^{2}}  
+f_{V}\tilde{g}_{V}\frac{M_{V}^{2}}{f_{\pi}^2}\frac{Q^2}{M_{V}^{2}}\frac
{Q^2}{M_{V}^2 +Q^2}\,. 
\ee  The first term $1$ is the  $\cO(p^2)$ contribution; the second term comes
from the $\tilde{L}_{9}$--coupling; the last one from integrating out the
vector field $V$ from the $\cO(p^3)$ interaction terms. Recall that
$\tilde{L}_{9}=\frac{1}{2}f_{V}\tilde{g}_{V}$, with $f_{V}$ and
$\tilde{g}_{V}$ defined in eqs.~(5.13) and (5.14).  Clearly,
\bea 
\tilde{F}(Q^2) & = &
1-f_{V}\tilde{g}_{V}\frac{M_{V}^{2}}{f_{\pi}^2}\frac{Q^2}{M_{V}^2}
\left(1-\frac{Q^2}{M_{V}^2+Q^2}\right) \\
 & = & 1-\frac{1+g_{A}}{2}\frac{Q^2}{M_{V}^2+Q^2}\,.
\eea  The problem with this form factor is that it does not satisfy the QCD
requirement that asymptotically in $Q^2$ it should fall as
$\frac{1}{Q^2}$. It does not have either the expected LMD behaviour
\be F(Q^2)=\frac{M_{V}^{2}}{M_{V}^2+Q^2}\,.
\ee  Notice that here we are talking about an on--shell form factor (the pions
are on--shell), and its imaginary part is supposed to have one state only,
the lowest vector meson state. There is no perturbative on--shell continuum
here.  In order to implement the correct large--$Q^2$ behaviour we are then
forced -once more- to introduce a set of an infinite number of local
operators in external fields. These are operators of the {\it second class
type} discussed in section~3.2\,. Their contribution to
$F(Q^2)$, let us call it $\tilde{f}(Q^2)$, when re--summed must give:
\be
\tilde{f}(Q^2)=-\frac{1-g_{A}}{2}\left\{\frac{Q^2}{M_{V}^2}-\frac{Q^4}{
M_{V}^2 (M_{V}^2 +Q^2)}\right\}\,,
\ee 
for then, the overall form factor will be
\be 
F(Q^2)=1-\frac{1+g_{A}}{2}\frac{Q^2}{M_{V}^2+Q^2}
-\frac{1-g_{A}}{2}\frac{Q^2}{M_{V}^2+Q^2}=
\frac{M_{V}^2}{M_{V}^2+Q^2}\,,
\ee 
as wanted.
The overall effect of this infinite number of local operators
is a simultaneous modification  of the coupling constant
$\tilde{g}_{V}$ and the constant 
$\tilde{L}_{9}$ in the effective Lagrangian in (\ref{eq:Leff4}) as follows:
\bea 
\tilde{g}_{V} & = & \frac{1+g_{A}}{2}\sqrt{1-g_{A}}\frac{f_{\pi}}{M_{V}}\ra
\left(\frac{1+g_{A}}{2}+\frac{1-g_{A}}{2} \right)
\sqrt{1-g_{A}}\frac{f_{\pi}}{M_{V}} \\
 & = &
\sqrt{1-g_{A}}\frac{f_{\pi}}{M_{V}}\equiv g_{V}\,,
\eea
and
\be
\tilde{L}_{9}=\frac{N_c}{16\pi^2}\frac{1}{6}(1-g_{A}^2)\Gamma_{0}\ra
\frac{N_c}{16\pi^2}\frac{1}{3}(1-g_{A})\Gamma_{0}\equiv L_{9}\,.
\ee
A similar modification of the coupling constants $\tilde{L}_{1}$,
$\tilde{L}_{3}$ also has to be made, accordingly, so as to preserve
the relations in eq.~(\ref{eq:froissart}) which ensure the short--distance
{\it matching} for four--point functions, with the result 
\bea
\tilde{L}_{1}=\frac{N_c}{16\pi^2}\frac{1}{48}(1-g_{A}^2)^2\Gamma_{0}\ra
\frac{N_c}{16\pi^2}\frac{1}{12}(1-g_{A})^2 \Gamma_{0}\equiv L_{1}\,, \\
\tilde{L}_{3}=-6\tilde{L}_{1}+g_{A}L_{5}\ra -6L_{1}+g_{A}L_{5}\equiv
L_{3}\,.
\eea 
The new coupling constants preserve the symmetries
\bea
L_{1} & = & \frac{1}{8}g_{V}^2\,,\\
L_{3} & = & -6L_{1}+g_{A}L_{5}\,,\\
L_{9} & = & \frac{1}{2}f_{V}g_{V}\,,
\eea
and, now, they also satisfy the relation
\be\label{eq:fgpirelation}  
f_{V}g_{V}=\frac{f_{\pi}^2}{M_{V}^2}\,.
\ee 

Let us next consider the axial form factor $G_{A}(Q^2)$ in
$\pi\ra e\nu\gamma$. Again, the condition of saturation of the imaginary part
with one axial--vector state, and no subtractions in the dispersion relation
implies~\cite{EGLPR89}: $2f_{V}g_{V}-f_{V}^2=0\,.$ It turns out that now,
using eqs.~(\ref{eq:fVa}) and (\ref{eq:fgpirelation}), there is indeed a
solution which satisfies this constraint, namely: 
\be g_{A}=\frac{1}{2}\,.
\ee
The net effect is that the number of
independent input parameters has now been reduced from three to two.

\vspace*{0.5cm}

\subsection{The LMD Effective Lagrangian of Large--$N_c$ QCD}

\noi   We finally find that after correcting the initial ENJL--Lagrangian
{\it ansatz} for the unwanted non--confining $Q\bar{Q}$--discontinuities and
demanding the correct {\it matching} between the leading QCD short--distance
behaviour of the VPP and PVA three--point functions and the 
asymptotic behaviour of the corresponding hadronic form factors,
the  resulting low--energy effective Lagrangian which describes LMD in
\QCD, to $\cO(p^4)$ in the chiral expansion, has the simple form 
\bea\label{eq:Leff4final} {\cL}_{\eff} & = &  \frac{1}{4} f_{\pi}^2\, 
\left[\tr\left(D_{\mu}UD^{\mu}U^{\dagger}\right) + \tr \left( \chi U^{\dagger}
    + U^{\dagger} \chi \right)\right] +\mbox{\rm e}^2\,C\tr
(Q_{R}UQ_{L}U^{\dagger}) 
\nn \\ & & -\frac{1}{4}\tr\,[V_{\mu\nu}V^{\mu\nu}-2M_{V}^2 V_{\mu}V^{\mu}]
 \nn \\ & &  -\frac{1}{4}\tr\,[A_{\mu\nu}A^{\mu\nu}-4M_{V}^2 A_{\mu}A^{\mu}] 
\nn
\\ & &  -\frac{1}{4}\frac{f_{\pi}}{M_{V}}\,\tr\,\left(2
V_{\mu\nu}f_{+}^{\mu\nu}+i V_{\mu\nu}[\xi^{\mu},
\xi^{\nu}]+ A_{\mu\nu}f_{-}^{\mu\nu}\right) \nn \\ & & 
   + {L}_1 \left(\tr\, D_{\mu}U^{\dagger} D^{\mu}U\right)^2+2{L}_1
\tr\left( D_{\mu} U^{\dagger} D_{\nu}U\tr D^{\mu}U^{\dagger}D^{\nu}U\right)
\nn
\\ & &  +{L}_3
\tr\left( D_{\mu}U^{\dagger} D^{\mu}UD_{\nu}U^{\dagger} D^{\nu}U
\right) \nn \\ & & + L_5\tr\left[D_{\mu}U^{\dagger} D^{\mu}U
\left(\chi^{\dagger}U+U^{\dagger}\chi\right)\right] + 
L_8\tr\left(\chi^{\dagger}U \chi^{\dagger}U +\chi U^{\dagger}\chi
U^{\dagger}\right) \nn \\ & & 
 -i{L}_9 \tr\left(F^{\mu\nu}_R D_{\mu}U  D_{\nu}U^{\dagger} +F^{\mu\nu}_L
D_{\mu}U^{\dagger} D_{\nu}U\right)+L_{10} \tr\left( U^{\dagger} F^{\mu\nu}_{R}
U F_{L\mu\nu}\right) \nn \\ & & 
+H_{1}\tr\left(F_{R}^{\mu\nu}F_{R\mu\nu}+F_{L}^{\mu\nu}F_{L\mu\nu}\right)+
H_{2}\tr\left(\chi^{\dagger}\chi\right)\,,
\eea with, so far, two free parameters $M_Q$ and $\Gamma_0$ which are such
that:
\be\label{eq:fpiMV}
f_{\pi}^2=\frac{N_c}{16\pi^2}2M_{Q}^2\Gamma_{0}\quad\annd\quad M_{V}^2
=\frac{1}{2}M_{A}^2 =6M_{Q}^2\,.
\ee The constant $C$ is then fixed,
\be\label{eq:Cpi}
\frac{2e^2 C}{f_{\pi}^2}=\frac{\alpha}{\pi}9M_{Q}^2 \log 2\,,
\ee and the Gasser--Leutwyler $L_i$ constants are all proportional to
$\frac{N_c}{16\pi^2}\Gamma_{0}$, i.e. to
$3 \frac{f_{\pi}^2}{M_{V}^2}$:
\be\label{eq:gasserleut}
6L_{1}=3L_{2}=\frac{-8}{7}L_{3}=4L_{5}=8L_{8}=\frac{3}{4}L_{9}=-L_{10}=
\frac{N_c}{16\pi^2}\frac{1}{8}\Gamma_0\,.
\ee
Notice that the dependence on the regularization chosen in the ENJL--{\it
ansatz} only appears as an undetermined overall constant $\Gamma_{0}$.
Different regularizations will give different ``expressions'' for
$\Gamma_{0}$, but once the value of $\Gamma_{0}$ is traded for a physical
observable, all the predictions are regularization independent.

When restricted to the vector and axial--vector sector, the Lagrangian
in (\ref{eq:Leff4final}) is entirely analogous to the phenomenological
Lagrangians discussed in ref.~\cite{EGLPR89}. They are just different
formulations of the same physics which describes the LMD of the $0^{-}$,
$1^{-}$ and $1^{+}$ hadronic states in
\QCD.
The predictive power of the low--energy effective Lagrangian in
(\ref{eq:Leff4final}), as compared to various phenomenologically inspired
Lagrangian formulations of resonances discussed e.g. in
refs.~\cite{EGPR89,DRV89,DHW93,DP97} and references therein lies in the fact
that in our case the vector, axial--vector \underline{and} scalar couplings
are all correlated by one mass scale $M_Q$ and one dimensionless  parameter
$\Gamma_{0}$, which can be traded for $f_{\pi}$ and $M_{V}$. In order to
predict the
$\cO(p^4)$
$L_{i}$--constants e.g., the phenomenological approaches require
$f_{\pi}$, $M_{V}$, the scalar couplings~\cite{EGPR89} $c_{m}$, $c_{d}$, and
$M_{S}$ as a minimum of input parameters. Going from $\cO(p^4)$ to
$\cO(p^6)$ in a phenomenological approach requires many more new input
parameters, while in our case they are all fixed by the
same
$M_Q$ and
$\Gamma_{0}$ parameters, provided that the {\it matching} constraints to
short--distance QCD for the new Green's functions which will appear at 
$\cO(p^6)$ can be made consistently with the LMD $0^{-}\,,
1^{-}\,,0^{+}\,,1^{+}$ spectrum only. In fact, we have seen that at
$\cO(p^4)$, the {\it matching} constraints to short--distance QCD beyond two--point
functions have actually reduced the number of three free parameters in the
initial ENJL--Lagrangian  {\it ansatz} to two. We have not yet made an
exhaustive study of this important question beyond $\cO(p^4)$.

One way to quantify the predictive power of the LMD approximation to \QCD
which we have been considering is to fix the remaining two parameters
$M_{Q}$ and $\epsilon\equiv \frac{M_{Q}^2}{\Lambda_{\chi}^2}$ 
( or $\Gamma_{0}$)
from a least square fit to experimentally well known quantities and compare
them with the resulting predictions. The choice of low--energy constants is
the one in the first column of Table~1.

\newpage

\begin{center}
\noi  {{\bf Table~1.}  {\it Low--Energy Constants, Experimental Values and
Least Square Fit.}}\\
\vskip 1pc
\begin{tabular}{|c|c|c|c|c|} \hline \hline {\em Parameter} & {\em Equation} &
{\em Exp. Value} & {\em Fit Error} & {\em Predicted Value}
\\ \hline $L_{2}$ & (\ref{eq:gasserleut}) & $(1.4\pm 0.3)\times 10^{-3}$ &
$\pm 0.3\times 10^{-3}$ & $1.7\times 10^{-3}$ \\
$L_{3}$ & (\ref{eq:gasserleut}) & $(-3.5\pm 1.1)\times 10^{-3}$ & $\pm
1.1\times 10^{-3}$ &
$-4.4\times 10^{-3}$ \\
$L_{5}$ & (\ref{eq:gasserleut}) & $(1.4\pm 0.5)\times 10^{-3}$ & $\pm
0.5\times 10^{-3}$ &
$1.3\times 10^{-3}$ \\
$L_{8}$ & (\ref{eq:gasserleut}) &  $(0.9\pm 0.3)\times 10^{-3}$ &  $\pm
0.3\times 10^{-3}$ &
$0.64\times 10^{-3}$ \\
$L_{9}$ & (\ref{eq:gasserleut}) & $(6.78\pm 0.15)\times 10^{-3}$ & $\pm
0.15\times 10^{-3}$ & $6.8\times 10^{-3}$ \\
$L_{10}$ & (\ref{eq:gasserleut}) & $(-5.13\pm 0.19)\times 10^{-3}$ & $\pm
0.19\times 10^{-3}$ & $-5.1\times 10^{-3}$ \\  
\hline
$f_{\pi}$ & (\ref{eq:fpiMV}) & $(92.4\pm 0.3)\,\MeV$ & $\pm 9\,\MeV$ &
$(87\pm 3.5)\,\MeV$
\\
$\Delta m_{\pi}$ & (\ref{eq:Cpi}) &  $(4.5936\pm 0.0005)\,\MeV$ &
$\pm 0.5\,\MeV$ &
$(4.9\pm 0.4)\,\MeV$ \\
\hline
$M_V$ & (\ref{eq:fpiMV}) & $(768.5\pm 0.6)\,\MeV$ & $\pm 75\,\MeV$ &
$(748\pm 29)\,\MeV$
\\
$M_{A}$ & (\ref{eq:fpiMV}) & $(1230\pm 40)\,\MeV$ & $\pm 200\,\MeV$ &
$(1058\pm 42)\,\MeV$ \\
 \hline \hline
\end{tabular}
\end{center}
\vskip 1pc
\noi  The equations which define them in the text are given in the second
column. The third column quotes their present experimental values with their
errors~\footnote{We use the recent determination of $L_{10}$ and the value of
$L_{9}$ quoted in ref.~\cite{DGHS98}.}. The fourth column shows the errors
which we have taken for the fit. The errors for the input parameters with a
mass dimension are taken larger than the corresponding experimental error, to
allow for possible effects due to next--to--leading
$1/N_c$--corrections and chiral corrections. From the nine constraints in
Table~1 and with the two parameters $M_{Q}$ and
$\epsilon\equiv\frac{M_{Q}^2}{\Lambda_{\chi}^2}$ as free parameters we obtain
a fit with a $\chi^2 = 4.2$. The corresponding values of the fit parameters
are then
\be\label{eq:MQandepsilon}   
M_{Q}=(305\pm 12)\,\MeV\,,\quad\annd\quad
\epsilon =0.0706\pm 0.0031\,.
\ee The resulting values for the $L_{i}$--constants and the other parameters
are given in the last column. Typically, the errors for the predicted
$L_{i}$'s are less than $2\%$; but of course this ignores the systematic
errors of the approximation which we are adopting. Figure 8 below shows a plot
of the predicted
$L_{i}'s$ in eq.~(\ref{eq:gasserleut}) versus
$\Gamma_{0}$.

\vspace{0.3cm}
\centerline{\epsfbox{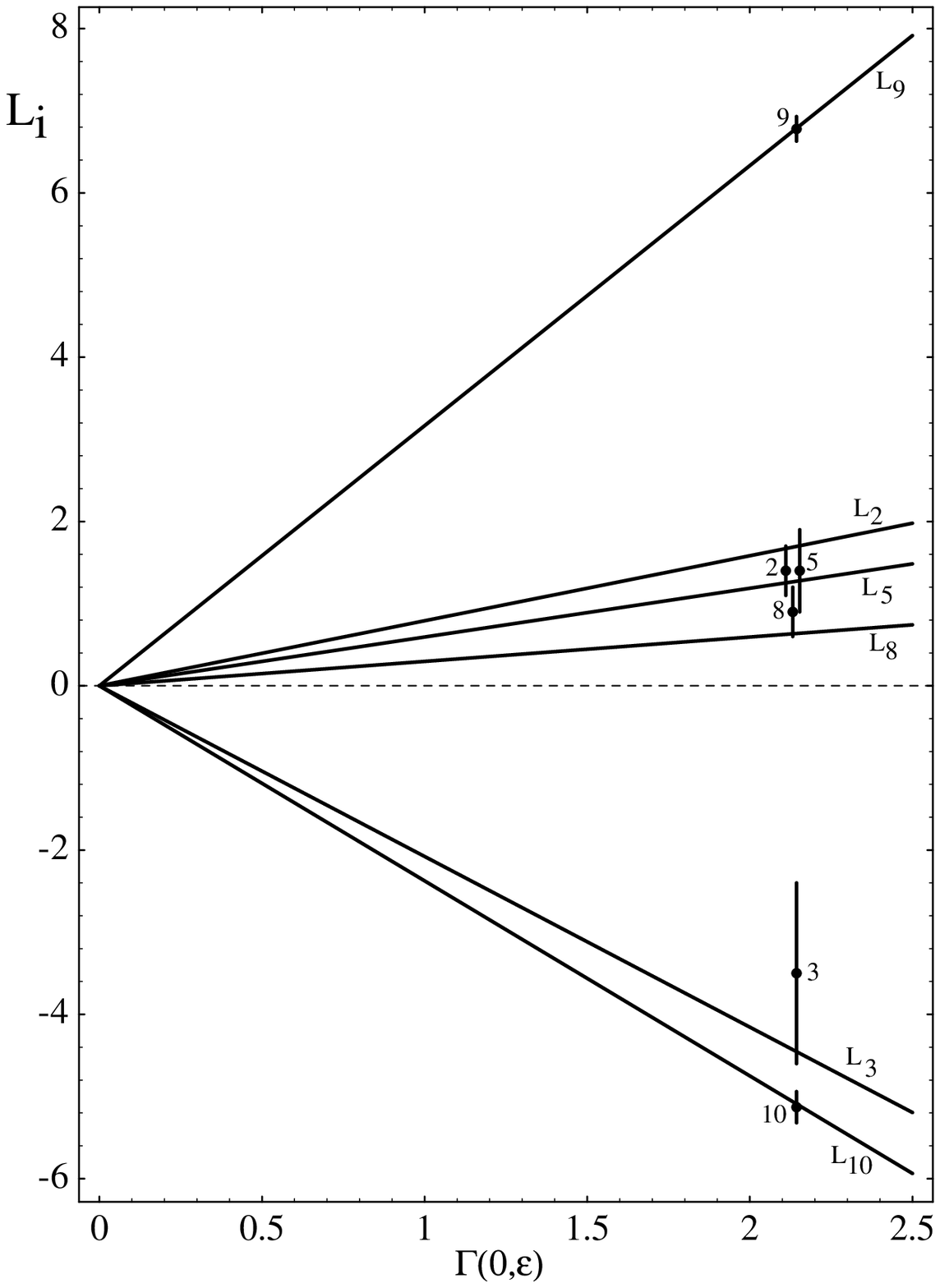}}
\vspace{0.2cm}
{\bf{Fig.~8}} {\it Plot of the predicted $L_{i}$ constants in
eq.~(\ref{eq:gasserleut}) versus $\Gamma_{0}$. The points with error bars
are the present phenomenological determination of these constants from
experiment.}

\vspace{0.3cm}
\noi 
It is
remarkable how well the phenomenological determination of these constants
intercept the predicted linear dependence at the \underline{same} value of
$\Gamma_{0}$. This is a clear improvement, both in the quality of the fit
(with one less parameter!) and in conceptual simplicity, to previous fits
made with the ENJL--Lagrangian~\cite{BBdeR93,Bi96} which were already quite
good.

All the other parameters which have appeared at various stages of the
discussion in the text are now fixed. For example the couplings of the initial
ENJL--Lagrangian are found to be  
\be\label{eq:GSGVLCHI} G_{S}=1.28\pm 0.01\,,\quad G_{V}=1.65\pm
0.04\,,\quad\annd\quad
\Lambda_{\chi}=(1150\pm 52)\,\MeV\,.
\ee 
Another parameter which has been discussed in the text is the onset of the QCD
continuum
$s_{0}$ in the vector and axial--vector spectral functions in
eqs.~(\ref{eq:largeN}) and (\ref{eq:axiallargeN}). This parameter
was fixed by the requirement that in the OPE of the corresponding
two--point functions there is no local operator of dimension $d=2$ in the
chiral limit. This leads to eq.~(\ref{eq:eigenv}) and hence, using the
determination for $f_{V}^2 M_{V}^2$ which follows from the effective
Lagrangian in (\ref{eq:Leff4final}) fixes the value of $s_{0}$ to 
\be\label{eq:matchings0}
s_{0}(1+\cdots)=6M_{Q}^2 \Gamma_{0}\simeq 1.20\,\GeV^2\,,
\ee 
where for simplicity
we have neglected gluonic perturbative corrections. Once the value of $s_{0}$
is fixed, the {\it matching} to higher $1/Q^2$--powers in the OPE fixes 
the corresponding combinations of condensates of gauge invariant local
operators. For example, {\it matching} the $\frac{1}{Q^4}$ powers in the
Adler function in the chiral limit, constrains the gluon condensate to
satisfy the sum rule
\be
\frac{1}{6}\frac{\als}{\pi}\langle G_{a}^{\mu\nu}G_{\mu\nu}^{a}\rangle=
-4f_{V}^2 M_{V}^4 +\frac{N_c}{16\pi^2}\frac{4}{3}s_{0}^{2}(1+\cdots)\,,
\ee which in terms of the two parameters $M_{Q}$ and $\Gamma_{0}$ and with
neglect of pQCD corrections results in
\be\label{eq:gluoncond}
\frac{\als}{\pi}\langle G_{a}^{\mu\nu}G_{\mu\nu}^{a}\rangle\simeq
\frac{N_c}{8\pi^2}(2\sqrt{3}M_{Q})^4 \,\Gamma_0 \left(\Gamma_0 -2\right)\,.
\ee
Numerically, this corresponds to a
value
\be
\als\langle G_{a}^{\mu\nu}G_{\mu\nu}^{a}\rangle\simeq 0.048\,\GeV^4\,, 
\ee
which is quite compatible with the phenomenological determinations (see
eq.~(\ref{eq:gluonc})).

Finally, the resulting shape of the Adler function, which was the starting
point of our analysis after all, is the dashed curve shown in Fig.~9 below. In
this curve, the gluonic corrections in the matching with the pQCD continuum
have been included. The agreement with the experimental shape (the full line)
is not as good as the VMD one with the experimental values of $M_{\rho}$ and
$f_{\rho}$ plotted in Fig.~4, but it is not bad.

\vspace{0.3cm}
\centerline{\epsfbox{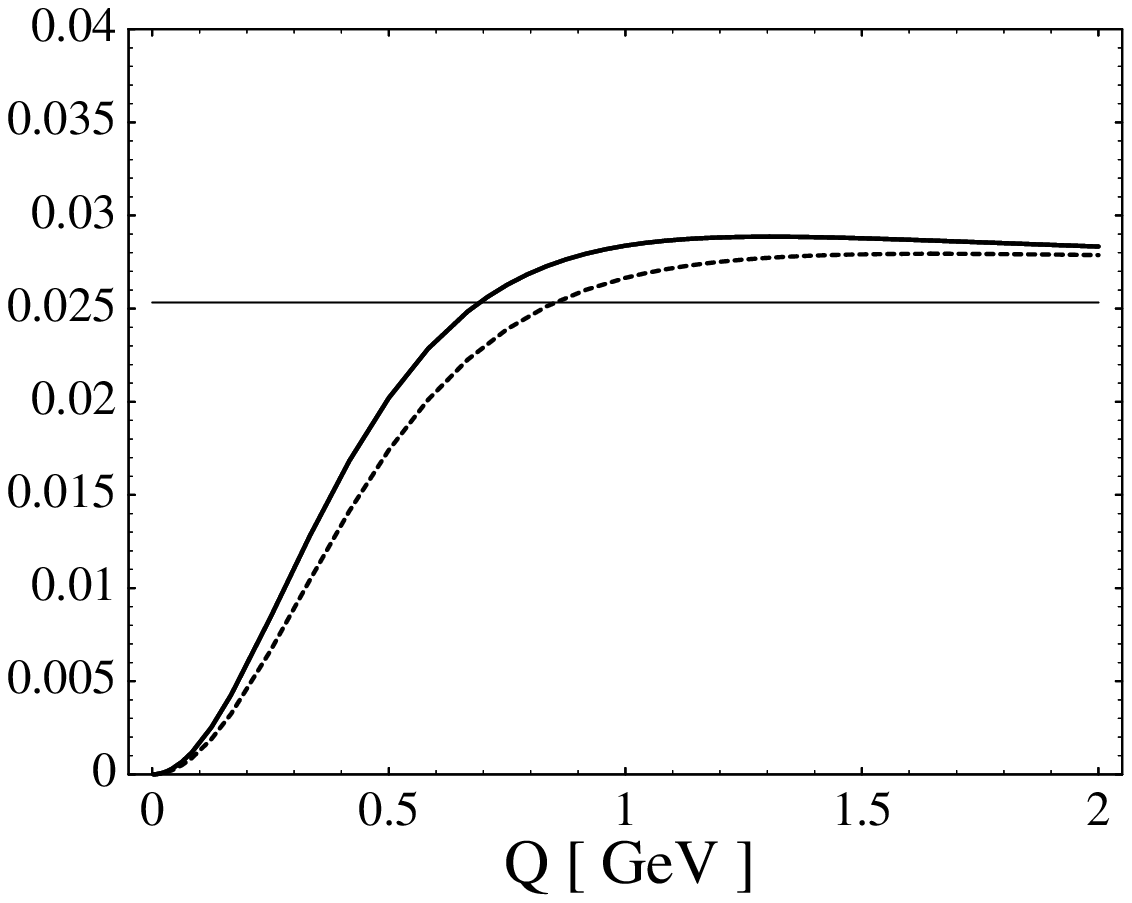}}
\vspace{0.2cm} {\bf{Fig.~9}} {\it The resulting Adler function of the LMD
approximation to large--$N_c$ QCD (dashed curve) compared to the
experimental data (full line).}

\vspace{0.3cm}

As discussed in ref.~\cite{KdeR97}, in the LMD approximation to large--$N_c$
QCD, and for given values of the masses of the lowest vector and
axial--vector states, all the local order parameters which appear in the OPE
of the
$\Pi_{LR}(Q^2)$ correlation function are fixed by the meson masses; in
particular the {\it matching} of the $\frac{1}{Q^6}$ powers leads to the sum
rule~\cite{KdeR97}
\be\label{eq:LRd6} f_{\pi}^2
M_{V}^2M_{A}^2=4\pi^2\frac{\als}{\pi}\left(1+\cO(\frac{\als^2}{\pi^2})
\right)\langle\bar{\psi}\psi\rangle^2\,,
\ee  which from eqs.~(\ref{eq:fpiMV}) can also be written as follows
\be\label{eq:3rdweinberg}
\frac{N_c}{4\pi^4}9M_{Q}^6 \Gamma(0,\epsilon)=\frac{\als}{\pi}
\left(1+\cO(\frac{\als^2}{\pi^2})
\right)\langle\bar{\psi}\psi\rangle^2\,.
\ee Using the values for the parameters $M_{Q}$ and $\epsilon$ which result
from the least square fit, i.e. the values in eq.~(\ref{eq:MQandepsilon}),
and the value of $\als$ at the {\it matching} scale in
eq.~(\ref{eq:matchings0}) we obtain
\be
\langle\bar{\psi}\psi\rangle
 = [(-288\pm 12)\,\MeV]^3\,,
\ee    on the high ball park of the phenomenological determinations (see
eq.~(\ref{eq:quarkc})) but certainly not an unreasonable value~\footnote{In
fact, the scale for the quark--condensate in eq.~(\ref{eq:quarkcond}) is
$s_{0}\simeq 1.2\,\GeV^2$, which is larger than the $1\,\GeV^2$ reference
scale used in eq.~(\ref{eq:quarkc}). The comparison at the same
$s_{0}$--scale is slightly better.}.  

One of the reasons why the ENJL--Lagrangian {\it ansatz} was chosen in the
first place is that above the critical value $G_{S}\ge 1$, it produces a
quark condensate, which in terms of $M_{Q}$ and $\epsilon$ is given
by~\cite{BBdeR93}
\be\label{eq:quarkcond}
\langle\bar{\psi}\psi\rangle
=\frac{-N_c}{16\pi^2}4M_{Q}^3\Gamma(-1,\epsilon)\,.
\ee So far, however,  we have not used this equation at all. {\it If} one is
willing to identify this quark condensate to the one of \QCD at a scale
$\mu\sim\Lac$; an interesting relation between the pQCD coupling constant and
non--perturbative parameters appears then when inserting
eq.~(\ref{eq:quarkcond}) in the r.h.s. of eq.~(\ref{eq:3rdweinberg}) and
using eq.~(\ref{eq:gapeq}) as well:
\be\label{eq:alphamatching}
\frac{\als N_c}{\pi}=\frac{9}{4}\frac{G_{S}^2}{G_{V}^2}
\frac{1}{\Gamma\left(0,\frac{M_{Q}^2}{\Lac^2}\right)}\,.
\ee 
This relation can be viewed as a {\it matching} constraint between pQCD above
the scale $\Lambda_{\chi}$ and the non--perturbative effective realization in
terms of hadronic degrees of freedom below $\Lac$. [Notice that
$\Gamma\left(0,M_{Q}^{2}/\Lac^2\right)=\log(\Lac^2/M_{Q}^{2})-
\gamma_{\mbox{\rm E}}+\cdots$.]  Numerically, the central value of the r.h.s.
of this equation which follows from the fit values in
eqs.~(\ref{eq:MQandepsilon}) and (\ref{eq:GSGVLCHI}) is 0.63, while
$\frac{\als N_c}{\pi}\sim
\frac{6}{11\log(\Lac/\Lmsb})\simeq 0.48$,  showing that the {\it matching},
at least in the vector and axial--vector channels, is not unreasonable. In
fact, one could use eq.~(\ref{eq:alphamatching}) to determine
$\epsilon\equiv\frac{M_{Q}^2}{\Lac^2}$, in which case the low--energy
parameters would all be constrained by just one mass scale.    

\vspace{0.3cm}
\section{Conclusions and Outlook}
\setcounter{equation}{0}
\label{sec:six}

We have shown that there exists a useful effective Lagrangian description of
a well defined approximation to QCD in the large--$N_c$ limit. This is the
approximation which restricts the hadronic spectrum in the channels with
$J^P$ quantum numbers $0^{-}$, $1^{-}$, $0^{+}$ and
$1^{+}$  to the lowest energy state and treats the rest of the narrow states
as a
\QCD~perturbative continuum, the onset of the continuum being fixed by
consistency constraints from the operator product expansion. The degrees of
freedom in the effective Lagrangian are then a nonet of pseudoscalar
Goldstone particles which are collected in a unitary matrix $U(x)$, and
nonets of vector fields $V(x)$, scalar fields $S(x)$ and axial--vector fields
$A(x)$ associated with the lowest energy states of the hadronic spectrum which
are retained. We have derived the effective Lagrangian by implementing
successive requirements on an ENJL--type Lagrangian (see eq.~(\ref{eq:njl}))
which we have chosen as the initial {\it ansatz}. The first requirement is to
eliminate the effects of {\it non--confining}
$Q\bar{Q}$ discontinuities by introducing an infinite number of appropriate
local operators with couplings which can be fixed in terms of the three
parameters of the starting ENJL--Lagrangian itself, i.e. $G_S$,
$G_V$ and the scale $\Lac$.  We have shown that the {\it matching} of the
two--point functions of this effective Lagrangian to their QCD
short--distance behaviour can be systematically implemented. In particular,
the 1st and 2nd Weinberg sum rules are automatically satisfied. The resulting
Adler function shown in Fig.~9, when confronted with the experimentally known
curve, shows a rather good interpolation of the intermediate region between
the two asymptotic regimes described by perturbative QCD (for the very
short--distances) and by chiral perturbation theory (for the very
long--distances).

For Green's functions beyond two--point functions, the removal of the {\it
non--confining}
$Q\bar{Q}$ discontinuities produced by the initial ENJL {\it ansatz} is
however not enough to guarantee in general the correct {\it matching} to the
leading QCD short--distance behaviour and further local operators have to be
included. We have discussed this explicitly in the case of the VPP and VPA
three--point functions, and shown that the {\it matching} with the QCD
short--distance leading behaviour which follows from the OPE restricts the
initial three free parameters of the ENJL--Lagrangian {\it ansatz} to just one
mass scale $M_{Q}^2$ and a dimensionless constant
$M_{Q}^2/\Lac^2$. The resulting low--energy Lagrangian in the vector and
axial--vector sector, and  to
$\cO(p^4)$ in the chiral expansion, coincides with the class of
phenomenological Lagrangians discussed in ref.~\cite{EGLPR89} which also have
two free parameters $f_{\pi}^2$ and $f_{\pi}^{2}/M_{\rho}^{2}$. In this
respect, this explains the relation to QCD of the phenomenological VMD
Lagrangians discussed in ref.~\cite{EGLPR89}: to
$\cO(p^4)$, they can be viewed as the effective low--energy Lagrangians of
the LMD approximation to \QCD. On the other hand, the fact that the resulting
low--energy Lagrangian coincides with the phenomenological VMD Lagrangians
discussed in ref.~\cite{EGLPR89} demystifies to a large extent the r{\^o}le
of the ENJL--Lagrangian itself  as a fundamental step in deriving the
low--energy effective Lagrangian of QCD. The ENJL--Lagrangian turns out to be
already a very good {\it ansatz} to describe in terms of quark fields degrees
of freedom the LMD approximation to
\QCD~and this is why it is already quite successful at the phenomenological
level; but, as we have seen, when the {\it non--confining} $Q\bar{Q}$
discontinuities are systematically removed, the phenomenological predictions
improve even more.  

There is an advantage, however, in starting with the ENJL--Lagrangian as an
{\it ansatz}, and that is that in this description of the LMD approximation to
\QCD, all the couplings to all orders in
$\chi$PT are clearly correlated to the same two free parameters. For example,
as shown in eqs.(\ref{eq:gasserleut}), the $L_5$ and $L_8$ constants, as well
as part of the contribution to the $L_3$ constant which result from scalar
exchanges, are now proportional to the same parameter
$\Gamma_{0}$, while in a purely phenomenological description in terms of
chiral effective Lagrangians which include resonances as discussed e.g. in
refs.~\cite{EGPR89,DRV89,DHW93,DP97} and references therein, these
low--energy constants require new phenomenological input. It is a very
interesting question to ask whether or not a systematic implementation of
short--distance OPE constraints on phenomenological Lagrangians leads to the
same description, to all orders in the chiral expansion, as the one starting
with the ENJL--Lagrangian {\it ansatz}. As we have said already, we have not
yet undertaken this study in an exhaustive way.     
   
Finally, we wish to insist on the phenomenological successes of the LMD
approximation to
\QCD, as demonstrated by the results collected in Table 1 and the plots in
Figs.~8 and 9. These results show that this is indeed a very good
approximation to full fledged QCD. It seems now worthwhile to apply it to the
calculation of couplings of
$\cO(p^6)$ and $\cO(e^2 p^2)$ as well. One can also reconsider non--leptonic
weak interactions in the light of this effective Lagrangian framework, where
we expect a much better {\it matching} between the long--distance evaluation
of matrix elements of four--quark operators and the short--distance pQCD
logarithmic dependence of the Wilson coefficients than e.g. in the cut--off
calculations of refs.~\cite{BBG87,Betal98} or the ENJL--calculations reported
in ref.~\cite{BP97}.

\vspace*{7mm} {\large{\bf Acknowledgments}}
\vspace*{3 mm}

\noi  We are grateful to J.~Bijnens, M.~Knecht and A.~Pich for very helpful
discussions at various stages of this work.

This work has been supported in part by TMR, EC-Contract No.
ERBFMRX-CT980169 (EURODA$\phi$NE). The work of S.~Peris has also been
partially supported by the research project CICYT-AEN95-0882.


\end{document}